\documentclass[12pt,english,12pt,sort&compress]{article}
\usepackage[T1]{fontenc}
\usepackage[latin9]{inputenc}
\usepackage{amsmath}
\usepackage{amssymb}
\usepackage{graphicx}
\usepackage{esint}
\usepackage[numbers]{natbib}

\makeatletter
\usepackage{times}\usepackage{epstopdf}

\usepackage{babel}

\usepackage{babel}

\makeatother

\usepackage{babel}
\begin{document}

\title{Coupled motion of asymmetrical tilt grain boundaries: molecular dynamics
and phase field crystal simulations}

\author{Z. T. Trautt,$^{a}$ A. Adland,$^{b}$ A. Karma$^{b}$ and Y. Mishin$^{a}$,%
\thanks{Corresponding author: ymishin@gmu.edu%
}}

\maketitle
\noindent $^{a}$ Department of Physics and Astronomy, MSN 3F3, George
Mason University, Fairfax, VA 22030, USA\\
$^{b}$ Department of Physics and Center for Interdisciplinary Research
on Complex Systems, Northeastern University, Boston, MA 02115, USA
\begin{abstract}
Previous simulation and experimental studies have shown that some
grain boundaries (GBs) can couple to applied shear stresses and be
moved by them, producing shear deformation of the lattice traversed
by their motion. While this coupling effect has been well confirmed
for symmetrical tilt GBs, little is known about the coupling ability
of asymmetrical boundaries. In this work we apply a combination of
molecular dynamics and phase field crystal simulations to investigate
stress-driven motion of asymmetrical GBs between cubic crystals over
the entire range of inclination angles. Our main findings are that
the coupling effect exists for most of the asymmetrical GBs and that
the coupling factor exhibits a non-trivial dependence on both the
misorientation and inclination angles. This dependence is characterized
by a discontinuous change of sign of the coupling factor, which reflects
a transition between two different coupling modes over a narrow range
of angles. Importantly, the magnitude of the coupling factor becomes
large or divergent within this transition region, thereby giving rise
to a sliding-like behavior. Our results are interpreted in terms of
a diagram presenting the domains of existence of the two coupling
modes and the transition region between them in the plane of misorientation
and inclination angles. The simulations reveal some of the dislocation
mechanisms responsible for the motion of asymmetrical tilt GBs. The
results of this study compare favorably with existing experimental
measurements and provide a theoretical ground for the design of future
experiments.

\noindent \textbf{Keywords:} Molecular dynamics, phase-field crystal,
grain boundary, shear stress
\end{abstract}

\section{Introduction\label{sec:Introduction}}

Recent research has led to the recognition that many grain boundaries
(GBs) in crystalline materials can couple to applied shear stresses
and are moved by them in a manner similar to dislocation glide \citep{Cahn04a,Suzuki05b,Cahn06a,Cahn06b,Mishin2010a,Molodov2011,Molodov-2011a,Molodov-2011b,Syed2012}.
During this ``coupled'' GB motion, the boundary produces shear deformation
of the lattice and causes relative translation of the grains parallel
to the GB plane. The coupling effect is characterized by a coupling
factor $\beta$ defined as the ratio of the tangential grain translation
velocity $v_{||}$ to the normal GB velocity $v_{n}$: $\beta=v_{||}/v_{n}$.
Coupling is considered perfect if $\beta$ is a geometrically determined
number depending only on crystallographic characteristics of the boundary
and not on its velocity or the driving stress.

Initially observed experimentally for low-angle GBs in Zn \citep{Li53,Bainbridge54},
the coupling effect has recently been demonstrated for many high-angle
GBs in a number of metallic and non-metallic materials \citep{Winning01,Winning02,Yoshida,Molodov07b,Legros08,Mompiou09a,Molodov09a,Gorkaya2009,Gorkaya2010,Molodov-2011a,Molodov-2011b,Molodov2011,Syed2012}.
The coupling can be responsible for the stress-induced grain growth
in nano-crystalline materials \citep{Monk06a,Zhu07a,Hemker07a,Legros08,Gianola08a}
and can influence the nucleation of new grains during recrystallization
\citep{Cahn09a}. Molecular dynamics (MD) simulations have been effective
in providing insights in the atomic mechanisms, geometric rules and
dynamics of coupled GB motion \citep{Suzuki05b,Cahn06a,Cahn06b,Mishin07a,Ivanov08a,Trautt-2012,Elsener2009,Wan2009,Zhang08,Monk06a}.
It has also been found that the phase field crystal (PFC) methodology
developed over the recent years \citep{Elder2002,Elder2004,Berry2006,Stefanovic2006,Wu-2007,Elder2007,Berry08,Mellenthin08,Stefanovic2009,Wu2010a,Jaatinen-2010,Spatschek2010,Olmsted2011,Adland-2012}
is well capable of reproducing equilibrium and non-equilibrium GB
properties. In particular, it predicts reasonable values of GB energies
\citep{Jaatinen-2010} and has been able to reproduce the phenomenon
of GB premelting \citep{Berry08,Mellenthin08} as well as non-trivial
structural transitions at GBs at high homologous temperatures \citep{Olmsted2011}.
In addition, it has been shown to reproduce the coupling effect for
both two-dimensional (2D) \citep{Spatschek2010} and 3D symmetric
tilt boundaries \citep{Adland-2012}.

Most of the experiments as well as simulations conducted so far have
been focused on symmetrical tilt GBs. The less-studied case of asymmetrical
tilt boundaries is more complex but poses new interesting questions.
For example, low-angle symmetrical tilt GBs are known to move by collective
dislocation glide in parallel slip planes \citep{Read50a}. By contrast,
low-angle asymmetrical GBs are composed of at least two different
types of dislocations. One would expect that the dislocations gliding
in intersecting slip planes could block each other and prevent the
coupled motion. In fact, the impossibility of coupled motion of asymmetrical
boundaries was suggested in the classical paper by Read and Shockley
\citep{Read50a}. Yet, recent MD simulations \citep{Taylor07,Zhang08,Zhang2009b},
bicrystal experiments \citep{Molodov2011,Syed2012} and the observation
of coupled GB motion in polycrystalline materials suggest that this
may not be the case. Even less is known about geometric rules of coupling
or migration mechanisms of large-angle asymmetrical GBs.

Based on purely geometric considerations, it was suggested that deviations
of the GB plane from symmetric inclinations should preserve the coupled
motion for both low and high-angle GBs \citep{Cahn06b}. Furthermore,
the coupling factor was predicted to be independent of the inclination
angle as long as the coupling mode remains the same (see discussion
of the multiplicity of coupling modes below) \citep{Cahn06b}. The
MD simulations \citep{Taylor07} of {[}001{]} tilt GBs in copper with
the tilt angle $\theta=18.92^{\circ}$ and several different inclinations
showed that $\beta$ did vary with the inclination angle within 10
to 20\%. However, the boundary remained coupled over the entire angular
range of inclinations as predicted. On the other hand, for large-angle
boundaries with $\theta=36.87^{\circ}$, two symmetrical and two asymmetrical
boundaries were tested and the coupling factors were found to be different
in both magnitude and sign (two $\beta$'s were positive and two negative)
\citep{Taylor07}. Zhang et al.~\citep{Zhang08} studied stress-driven
motion of asymmetrical {[}001{]} tilt GBs in Ni with $\theta=36.87^{\circ}$
and $\theta=33.36^{\circ}$. They observed coupled motion (with varying
sign of $\beta$) in some cases and sliding in other cases. The inconclusive
and often conflicting results of the previous studies point to the
need for a more detailed and systematic analysis of coupling of asymmetrical
GBs.

In this paper we report on simulations of stress-driven motion of
a large set of asymmetrical {[}001{]} tilt GBs by applying two complementary
methodologies: MD and PFC. The MD simulations are conducted for specific
materials (FCC copper and aluminum) and provide quantitative information
about the mechanical stresses needed for driving the GB motion at
different temperatures. The MD simulations are also well suited for
studying atomic-level mechanisms of GB migration by examining atomic
trajectories. A weakness of the MD approach is that the time scale
is limited to tens of nanoseconds, preventing access to diffusion-controlled
processes such as dislocation climb.

The PFC methodology overcomes, in principle, the latter limitation
by permitting simulations on diffusive time scales, thereby describing
both dislocation glide and climb \citep{Berry2006}. This gives us
the opportunity to get a glimpse into the possible GB behavior in
the long-time regime which is currently unaccessible by MD simulations.
However, one limitation of the PFC method is that the number of peaks
of crystal density waves does not directly correspond to the number
of atoms, which is not generally conserved. While this permits a description
of dislocation climb phenomenologically, it remains unclear how to
define and control the vacancy concentration in the PFC model. Theoretical
attempts have been made to incorporate vacancies explicitly in the
PFC model, but they do not fully resolve this issue since climb remains
possible even for a vanishing vacancy concentration \citep{Chan-2009}.
As a consequence, even though the geometrical aspects of the glide-mediated
conservative GB motion are well modeled by PFC quantitatively, the
description of non-conservative climb-mediated motion remains largely
qualitative. 

Despite this limitation, we find here that the MD and PFC approaches
reveal remarkably similar coupling behaviors as a function of the
misorientation and inclination angles, including the existence of
a narrow transition region between two coupling modes and a sliding-like
behavior in this transition region. Hence, a comparison of the MD
and PFC simulations helps us shed light on aspects of GB behavior
that do not depend sensitively on detailed atomic mechanisms. Finally,
in order to be able to explore efficiently the entire parameter space
of the misorientation and inclination angles, we restrict our PFC
simulations to 2D GBs between square lattices. We make use of the
PFC model of Ref.~\citep{Wu2010a}, which energetically favors the
square and FCC lattices in 2D and 3D, respectively. Even though the
2D square geometry is simpler than the 3D FCC structure studied in
the MD simulations, we are able to quantitatively relate the Burgers
vector character of the GBs in the PFC and MD simulations owing to
the fact that the square lattice rotated by $45{}^{\circ}$ is identical
to one plane of the FCC lattice. Hence, the choice of the 2D geometry
is not highly restrictive for the purposes of the present study.

\section{Simulation geometry\label{sec:symmetry}}

The geometry of asymmetrical {[}001{]} tilt GBs between FCC crystals
is illustrated in Fig.~\ref{fig:Simulation-Geometry}. Because the
orientations of the GB plane and the tilt axis are fixed, the system
has two geometric degrees of freedom. These degrees of freedom can
be associated with the tilt angle $\theta$ and the inclination angle
$\phi$. The tilt angle is defined as the misorientation angle between
the $[001]$ directions in the grains, with $\theta=0$ corresponding
to a single crystal and $\theta\geq0$ to the $[001]_{1}$ axis rotated
in the counter-clockwise direction relative to $[001]_{2}$. The inclination
angle $\phi$ is defined as the angle between the GB plane and the
internal bisector between the $[001]$ directions in the grains. We
take $\phi\geq0$ if the bisector is rotated in the counter-clockwise
direction relative to the GB plane. The case of $\phi=0$ with $\theta\neq0$
corresponds to a symmetrical tilt GB.

Due to the fourfold symmetry of the FCC lattice, all distinct GB structures
can be found in the angular domain $\left\{ 0\leq\theta<\pi/4,\;0\leq\phi<\pi/4\right\} $.
However, we are interested in not only the GB structures but also
their orientations relative to the laboratory coordinate system $xyz$
(Fig.~\ref{fig:Simulation-Geometry}). In order to include all possible
orientations of the GB structures, we consider an expanded domain
$\left\{ 0\leq\theta<\pi/2,\;-\pi/4\leq\phi<\pi/4\right\} $.

Symmetry analysis gives the following orientation relationships between
the GB structures: 
\begin{equation}
\left(\theta,\phi\right)\overset{m_{y}}{\rightleftharpoons}\left(\theta,-\phi\right),\label{eq:1}
\end{equation}
 
\begin{equation}
\left(\theta,\phi\right)\overset{m_{x}}{\rightleftharpoons}\left(90^{\circ}-\theta,45^{\circ}-\phi\right),\;\phi\geq0,\label{eq:2}
\end{equation}
 
\begin{equation}
\left(\theta,\phi\right)\overset{m_{x}}{\rightleftharpoons}\left(90^{\circ}-\theta,-45^{\circ}-\phi\right),\;\phi\leq0.\label{eq:3}
\end{equation}
 Eq.~(\ref{eq:1}) states that the GB structures with the same $\theta$
but opposite signs of $\phi$ are mirror reflections of each other
across the GB plane $(x,z)$. Eqs.~(\ref{eq:2}) and (\ref{eq:3})
state that replacement of the angles $\theta$ and $\phi$ by the
complementary angles $\left(90^{\circ}-\theta\right)$ and $\left(\pm45^{\circ}-\phi\right)$
reflects the GB structure across the mirror plane $m_{x}$ normal
to the $x$-axis. In particular, the inclination angles $\phi=\pm45^{\circ}$
for a given $\theta$ produce identical symmetrical tilt boundaries
which are $m_{x}$-reflections of the symmetrical boundary with $\phi=0$
and the tilt angle $\left(90^{\circ}-\theta\right)$. As will be discussed
later, these symmetry relations impose certain conditions on the response
of the GBs to applied shear stresses. The symmetry relations (\ref{eq:1})-(\ref{eq:3})
also apply to the 2D GBs modeled by the PFC method with an appropriate
choice of the lattice unit cell.

\section{Multiplicity of coupling modes\label{sec:Multiplicity}}

When a GB executes coupled motion, its dislocation content shears
the lattice swept by its motion and simultaneously rotates the lattice
to align it with the lattice of the growing grain. This combination
of shear and rotation produces relative translation of the two grains
parallel to the GB plane by the amount $\beta L$, where $L$ is the
normal GB displacement. As was discussed in previous work \citep{Suzuki05b,Cahn06a,Cahn06b},
the coupling factor $\beta$ is a multi-valued function of the crystallographic
angles of the boundary. There are two ways to understand the origin
the multiplicity of coupling factor.

One way is to recognize that the shear deformation produced by the
GB depends on its dislocation content, which, for a general GB, is
defined by the Frank-Bilby equation \citep{Balluffi95}. The latter
is known to have multiple solutions due to the crystal symmetry, leading
to the multiplicity of dislocation content of the GB. Different dislocation
contents produce different deformations of the lattice, which is manifested
in different coupling factors observed during the boundary motion.
This results in the existence of multiple coupling modes of the same
GB, each corresponding to a different solution of the Frank-Bilby
equation.

Another interpretation of the coupled motion focuses on the lattice
rotation step. Consider a tilt GB as an example. To ensure continuity
of the lattice of the growing grain, the receding lattice must be
rotated around the tilt axis by the angle $\pm\theta$ (the sign depends
on the direction of GB motion). However, if the lattice possesses
$n$-fold rotation symmetry around the tilt axis, then rotations by
the angles $\pm\theta+(2\pi/n)$, $\pm\theta+2(2\pi/n)$, etc., also
produce physically identical states of the growing grain. But these
different rotation angles lead to different relative translations
of the grains and thus different coupling factors.

For the {[}001{]} tilt axis in a cubic material, the fourfold symmetry
generates four possible coupling modes, with the coupling factors
$\beta=2\tan\left(\theta/2+\pi k/4\right)$, $k=0,1,2,3$. In reality,
only two of them, corresponding to the smallest magnitude of $\beta$,
have been observed in MD simulations \citep{Suzuki05b,Cahn06a,Cahn06b}
and experiments \citep{Molodov07b,Molodov09a,Gorkaya2009,Gorkaya2010,Molodov2011}.
These two modes are referred to as $\left\langle 100\right\rangle $
and $\left\langle 110\right\rangle $ type and are characterized by
the coupling factors 
\begin{equation}
\beta_{\left\langle 100\right\rangle }=2\tan\left(\dfrac{\theta}{2}\right)\label{eq:mode1}
\end{equation}
 and 
\begin{equation}
\beta_{\left\langle 110\right\rangle }=2\tan\left(\dfrac{\theta}{2}-\dfrac{\pi}{4}\right),\label{eq:mode2}
\end{equation}
respectively. Note that these two coupling factors have different
signs and describe GB motion in opposite directions in response to
the same shear. For symmetrical tilt boundaries, the coupling factors
obtained by simulations \citep{Suzuki05b,Cahn06a,Cahn06b} and experiments
\citep{Molodov07b,Molodov09a,Gorkaya2009,Gorkaya2010,Molodov2011}
accurately follow the $\left\langle 100\right\rangle $ mode for angles
$\theta<36^{\circ}$ and the $\left\langle 110\right\rangle $ mode
for angles $\theta>36^{\circ}$. At the critical angle of approximately
$36^{\circ}$, $\beta$ abruptly changes from one mode to the other
and can exhibit a ``dual behavior'' in which the boundary switches
back and forth between the two modes \citep{Suzuki05b,Cahn06a,Cahn06b}.
It should be noted that this switching angle is not prescribed by
any symmetry requirements and its exact value may depend on the material
and/or the crystal structure. In particular, in a previous MD study
of Cu \citep{Cahn06b}, an examination of the gamma-surfaces for different
coupling modes (Fig.~24 in \citep{Cahn06b}) revealed that the slip
responsible for the $\left\langle 100\right\rangle $ mode is more
difficult than the slip corresponding to the $\left\langle 110\right\rangle $
mode. The lower Peierls-Nabbaro (PN) barrier for slip in the $\left\langle 110\right\rangle $
mode is consistent with the fact that this mode spans a larger range
of misorientation, and therefore that the switching angle is less
than $45^{\circ}$.

We now discuss the coupling factors of asymmetrical tilt GBs. Suppose,
as suggested in \citep{Suzuki05b,Cahn06a,Cahn06b}, the coupling factor
for a given mode is independent of the inclination angle. Then one
can hypothesize that the expected angle dependence of $\beta$ would
look as shown schematically in Fig.~\ref{fig:Schematic-modes}. Only
the $\left\langle 100\right\rangle $ and $\left\langle 110\right\rangle $
modes are considered and their coupling factors are represented by
two surfaces. The cut between the surfaces corresponds to the discontinuous
transition between the modes accompanied by a reversal of sign. The
exact shape of this cut is unknown and it is drawn in this figure
with the only requirement that it respect the symmetry relations (\ref{eq:1})-(\ref{eq:3}).
When applying these relations, it was taken into account that the
reflection of the GB structure across its plane $(x,z)$ does not
affect the coupling factor. By contrast, reflection of the GB structure
across the mirror plane $m_{x}$ normal to the $x$-axis reverses
the sign of $\beta$. In particular, if for symmetrical boundaries
with $\phi=0$ the mode switching occurs at a tilt angle of $\theta=36^{\circ}$,
then for symmetrical boundaries with $\phi=\pm45^{\circ}$ it occurs
at $\theta=90^{\circ}-36^{\circ}=54^{\circ}$. Accordingly, all boundaries
with $\theta<36^{\circ}$ are expected to have the same positive coupling
factor, given by Eq.~(\ref{eq:mode1}), regardless of $\phi$. Likewise,
all boundaries with $\theta>54^{\circ}$ are expected to have the
same negative coupling factor given by Eq.~(\ref{eq:mode2}) regardless
of $\phi$. For boundaries with $36^{\circ}<\theta<54^{\circ}$, $\beta$
is positive at and near $\phi=0$ but is expected to switch to negative
values when approaching $\phi=45^{\circ}$ and $\phi=-45^{\circ}$.
Although the exact switching angles may depend on the material, this
analysis gives rather definitive qualitative predictions that can
be tested by simulations.

It should be emphasized again that the shape of the cut between the
surfaces in the putative diagram of Fig.~\ref{fig:Schematic-modes},
and even its existence, are at this point hypothetical. It is the
goal of this paper to test this diagram by simulations. Our simulation
results reported later in this paper will reveal that this diagram
is qualitatively correct in identifying the overall shape of regions
of opposite signs of $\beta$. However, we will see that it fails
to capture the fact that the magnitude of $\beta$ depends on the
inclination angle and can become very large near the discontinuous
boundary between the two coupling modes.

\section{Methodology of atomistic simulations\label{sec:Methodology-MD}}

Atomic interactions were modeled using embedded-atom method (EAM)
potentials fit to experimental and first-principle data for Cu and
Al \citep{Mishin01,Mishin99b}. Both potentials accurately reproduce
physical properties of these metals that are important in the context
of this study. In particular, they predict accurate values of the
elastic constants and stacking fault energies as well as dislocation
core structures. The melting points predicted by these potentials
are $T_{m}=1327$ K for Cu and $T_{m}=1040$ K for Al (the experimental
values are 1357 K and 933 K, respectively). It should be noted that
Al and Cu have significantly different stacking-fault energies and
thus different dislocation core splittings and mobilities. The inclusion
of both metals under the same methodology was intended to lend this
work more generality and facilitate comparison with possible future
experiments.

The GBs were created by constructing two separate crystals and joining
them along a plane normal to the $y$-direction (Fig.~\ref{fig:Simulation-Geometry}).
Periodic boundary conditions were imposed in the $x$- and $z$-directions
parallel to the GB plane. To achieve commensurability of lattice plane
periodicities in the grains in the $x$- and $z$-directions, the
angles $\theta$ and $\phi$ were chosen so that to create a coincident
site lattice (CSL) and align one of the CSL planes with the GB plane.
The dimensions of the simulation block in the $x$- and $z$-directions
comprised integral numbers of CSL periods, which completely eliminated
coherency strains. In the $y$-direction, the grains were terminated
at free surfaces. Several atomic layers near each surface were exempt
from the MD process and were used only to control the boundary conditions.
Namely, atoms in surface layer 2 were fixed in their perfect lattice
positions during all simulations, whereas atoms in surface layer 1
were fixed only in the $y$- and $z$-directions and were displaced
with the same constant velocity in the $x$-direction. All other atoms
were dynamic. Approximate dimensions of the simulation block were
$L_{X}\approx100\textrm{-\ensuremath{200}\AA}$, $L_{Y}\approx370\textrm{\AA}$
and $L_{Z}\approx37\textrm{\AA}$ and the total number of atoms was
$1\times10^{5}$ to $2\times10^{5}$.

The equilibrium GB structure was obtained by the energy minimization
procedure described in \citep{Suzuki03a}. Prior to the MD simulation,
the block was uniformly expanded by the thermal expansion factor at
the chosen simulation temperature. This expansion was intended to
eliminate thermal stresses inside the grains. The thermal expansion
factors for EAM Cu and Al were known from separate calculations \citep{Mishin01,Ivanov08a}.
The MD simulations were performed in the canonical (NVT) ensemble
with temperature controlled by a Nose-Hoover thermostat. A 2 femtosecond
time integration step was used throughout this study. After temperature
reached the target value, the GB was equilibrated by an isothermal
anneal for a few hundred picoseconds. The equilibration was followed
by a production run in which the surface layer 1 was moved parallel
to the $x$-direction with the speed of $v=0.5$ m/s, imposing a shear
strain rate of about $10^{7}$ s$^{-1}$. Shears in both positive
and negative $x$-directions were implemented for each boundary. The
production runs took up to 50 nanoseconds but were often terminated
earlier when the GB reached one of the surface layers (Fig.~\ref{fig:Simulation-Geometry}).
Multiple snapshots of the simulation block containing the coordinates
of all atoms, their energies and other relevant information were saved
during the simulations.

The following technique was applied for tracking the GB motion. First,
an orientation parameter, $\lambda$, was assigned to each individual
atom in a given snapshot. To this end, vector positions $\mathbf{r}_{ij}$
of $n\approx50$ nearest neighbors of a chosen atom $i$ were compared
with positions $\hat{\mathbf{r}}_{m}(\alpha)$ of 12 first-nearest
neighbors of an atom in the perfect FCC lattice rotated by an angle
$\alpha$ around {[}001{]}:

\begin{equation}
\psi_{i}(\alpha)=\sum_{j=1}^{n}\sum_{m=1}^{12}\sum_{k=1}^{3}\exp\left[-\dfrac{(r_{ijk}-\hat{r}_{mk}(\alpha))^{2}}{a^{2}}\right].\label{eq:orientation-1}
\end{equation}
Here $a$ is the equilibrium lattice constant and index $k$ runs
over three Cartesian components of a vector. The quantity $\psi_{i}(\alpha)$
was calculated for two angles $\alpha$ corresponding to the chosen
rotations of the grains relative to the coordinate system. The angle
with the larger $\psi_{i}(\alpha)$ provided a better match between
the actual and ideal lattice orientations and was assigned to atom
$i$ as its orientation parameter $\lambda$. Fig.~\ref{fig:Orientation-parameter}
illustrates this procedure by showing the atoms assigned to the grains
by the bright and dark colors.

After partitioning the atoms between the grains, the mean GB position
$h$ was calculated as $h=L_{Y}N/N_{0}$, where $N$ is the number
of atoms in grain 1 and $N_{0}$ is the total number of atoms in the
simulation block. The plots of $h$ versus time were used for calculation
of the coupling factor $\beta$. The first and last 15 $\textrm{\AA}$
of the plot were disregarded to eliminate the effects of the elastic
strain and interaction with the surface layers. The remaining part
of the plot was linearized by a least-mean-square fit and the slope
$S$ of the regression line was used to compute $\beta=v/S$.

Various visualization methods were applied for examining the structures
of moving GBs. In particular, for GBs composed of discrete dislocations
their slip traces contained information about the dislocation movements
and reactions. The deformation fields method proposed in \citep{Tucker2010}
was implemented and applied for visualization of dislocations and
their slip traces during the GB motion.

\section{Methodology of PFC simulations\label{sec:Methodology-PFC}}

The PFC model is generally formulated as an evolution equation for
the dimensionless crystal density field $\psi(\vec{r},t)$ defined
as the departure of the atomic number density $n(\vec{r},t)$ from
some reference value $n_{0}$ normalized by that value, $\psi(\vec{r},t)=(n(\vec{r},t)-n_{0})/n_{0}$.
Here PFC simulations are carried out in the grand canonical ensemble
with a constant chemical potential $\mu$, so that $\psi$ is a non-conserved
field. We use the dynamical formulation \citep{Stefanovic2009} 
\begin{equation}
\frac{\partial^{2}\psi}{\partial t^{2}}+\alpha\frac{\partial\psi}{\partial t}=-\frac{\delta F}{\delta\psi}+\mu,\label{PFCdyn}
\end{equation}
where the second order partial derivative with respect to time, $\partial^{2}\psi/\partial t^{2}$,
has the advantage of rapidly relaxing the elastic field via propagation
of wave-like modes that mimic phonons in a real solid. The first order
partial derivative with respect to time damps those modes and also
models diffusive processes. The total free-energy of the system is
expressed in terms of the functional 
\begin{equation}
F=\int d\vec{r}f+F_{ext},
\end{equation}
where $f$ is the free-energy density of the bulk crystal chosen to
have the form \citep{Wu2010a} 
\begin{equation}
f=\frac{\psi}{2}\left(-\epsilon+(\nabla^{2}+1)^{2}\left[(\nabla^{2}+Q_{1}^{2})^{2}\right]\right)\psi+\frac{\psi^{4}}{4},\label{feen}
\end{equation}
and $F_{ext}$ specified below models an ``external'' potential
used to shear the system. The form of $f$ couples two sets of crystal
density waves with different reciprocal lattice vectors, where $Q_{1}$
is the ratio of the magnitudes of those vectors. This form models
FCC lattices in 3D by coupling $[111]$ and $[200]$ reciprocal lattice
vectors with $Q_{1}=\sqrt{4/3}$, and square lattices in 2D by coupling
$[10]$ and $[11]$ reciprocal lattice vectors for the choice $Q_{1}=\sqrt{2}$,
which is adopted here.

As in the MD simulations, the GBs were created by constructing two
separate crystals and joining them along a plane normal to the $y$-direction
(Fig.~\ref{fig:Simulation-Geometry}). We use identical definitions
and sign conventions for the tilt and inclination angles as in the
MD simulations. Periodic boundary conditions were applied in the $x$-direction,
with the angles $\theta$ and $\phi$ and the dimension $L_{X}$ chosen
to be equal to an integral number of CSL periods in order to eliminate
coherency strains in the crystals. The dimension $L_{Y}$ was chosen
to be large enough to obtain GB behaviors that are independent of
$L_{Y}$, with typical values in the range of $100$ lattice spacings.

To enable comparison between the PFC and MD results, we replaced the
primitive square unit cell of the 2D PFC lattice with edge $a$ by
an expanded unit cell rotated by $45^{\circ}$ with edge $\sqrt{2}a$,
which contains 2 atoms (Fig.~\ref{fig:PFC-80degrees}). As a result,
the PFC structure becomes identical up to a scaling factor to one
of the (002) crystal planes of the 3D FCC structure. With this choice
of the unit cell, the Burgers vectors of the two types of dislocations
present in the MD and PFC lattices have the same Miller indices: $\langle100\rangle$
and $1/2\langle110\rangle$ in MD and $\langle10\rangle$ and $1/2\langle11\rangle$
in the PFC simulations, respectively. Henceforth, all crystallographic
indices related to the PFC structures will be given with respect to
the expanded unit cell. Fig.~\ref{fig:PFC-80degrees} illustrates
$1/2\langle11\rangle$ dislocations forming a low-angle symmetrical
tilt GB. The Burgers vector is given by the closure failure of the
Burgers circuit drawn around one of the dislocations.

Since the PFC method does not model a solid-vacuum interface, we cannot
use free surfaces as in the MD simulations to shear the bicrystal.
However, we can mimic free surfaces by choosing the chemical potential
to vary spatially over narrow strips near the top and bottom surfaces
of the crystal for the geometry depicted in Fig.~\ref{fig:Simulation-Geometry}.
This is accomplished by choosing the chemical potential $\mu$ to
vary spatially along the direction $y$ normal to the GB from a value
$\mu_{s}$ which favors the solid phase in most of the bulk of the
sample to a value $\mu_{l}$ which favors the liquid phase near the
top and bottom surfaces. With $y$ varying from $0$ to $L_{Y}$,
where $0$ corresponds to the bottom surface, the choice 
\begin{equation}
\mu=\mu_{l}+\frac{(\mu_{l}-\mu_{s})}{2}\left(\tanh\left[(y-L_{Y}+b)/\xi\right]-\tanh\left[(y-b)/\xi\right]\right)\label{muramp}
\end{equation}
melts the top and bottom surfaces a distance $b$ into the sample
if the parameter $\xi$ is chosen much smaller than $b$. To determine
the values of $\mu_{l}$ and $\mu_{s}$, we first compute the solid-liquid
phase diagram by a standard common tangent construction where the
free-energy of the solid phase is computed by expanding the crystal
density field in terms of the two sets of $[10]$ and $[11]$ density
waves as described in \citep{Wu2010a}. This construction yields the
value of the equilibrium chemical potential, $\mu_{E}$, for solid-liquid
coexistence. The solid (liquid) phase is favored for $\mu>\mu_{E}$
($\mu<\mu_{E}$) on the side of the phase diagram where $\mu_{E}$
is negative and the average solid density is larger than the liquid
density. We choose $\mu_{s}=0.9\mu_{E}$ and $\mu_{l}=1.1\mu_{E}$
for all simulations reported here. 

A shear is imposed by choosing 
\begin{eqnarray}
F_{ext} & = & \int dxdy\,\left[G(y-d)\left(\psi(x,y,t)-\psi_{0}(x-vt,y)\right)^{2}\right.\nonumber \\
 &  & +\left.G(y-L_{y}+d)\left(\psi(x,y,t)-\psi_{0}(x+vt,y)\right)^{2}\right],\label{shearPFC}
\end{eqnarray}
where $G(y)=\exp(-y^{2}/2\sigma^{2})/\sqrt{2\pi\sigma^{2}}$ is a
normalized Gaussian of width $\sigma$ and $\psi_{0}(x,y)$ corresponds
to the analytical expression for the equilibrium $\psi$ field for
a perfect single crystal approximated as a superposition of the $[10]$
and $[11]$ sets of crystal density waves (Eq.~(87) in \citep{Wu2010a}).
Since the dynamics tends to drive the system towards a minimum of
the grand potential $F-\mu\int d\vec{r}\psi$, the terms proportional
to $\left(\psi(x,y,t)-\psi_{0}(x\pm vt,y)\right)^{2}$ in the integrand
of $F_{ext}$ tend to entrain $\psi$ in a way that is equivalent
to pulling the crystal in opposite directions along $x$ at velocity
$\pm v$. The Gaussian kernels enforce that this pulling only takes
place in narrow strips of width $\sigma$ centered at a distance $d$
into the sample from its bottom and top surfaces. In general, $d$
needs to be chosen much smaller than $L_{y}$ and slightly larger
than the width $b$ of the surface melted layers so that the pulling
is applied to solid strips near the bottom and top surfaces that are
not melted.

All the simulations were carried out with $\alpha=0.1$ and $v=1.25\times10^{-4}$,
where $2v$ is the relative velocity of the two crystals (cf. Eq.~(\ref{shearPFC})).
A series of simulations of symmetrical tilt boundaries and all the
simulations of asymmetrical tilt boundaries were carried out with
$\epsilon=0.25$ for which $\mu_{E}=-1.33215$ (recall that $\mu_{s}=0.9\mu_{E}$
and $\mu_{l}=1.1\mu_{E}$ in Eq.~(\ref{muramp})). We also repeated
the series of simulations for symmetrical tilt boundaries with $\epsilon=0.05$
and thus $\mu_{E}=-0.593103$. This choice was motivated by the fact
that the discontinuous transition between the two different coupling
modes is expected from the previous MD work \citep{Cahn06b} reviewed
above to occur at a critical misorientation that is controlled by
the relative magnitudes of the PN barriers to dislocation motion in
different coupling modes. For the higher $\epsilon$ value ($\epsilon=0.25$),
the PN barrier turns out to be substantially larger for the $\langle10\rangle$
than $1/2\langle11\rangle$ dislocations \citep{Adland-2012b}. Since
low-angle symmetrical tilt boundaries with small $\theta$ are composed
of $\langle10\rangle$ dislocations, the coupled motion is expected
to switch from the $\langle10\rangle$ to the $\langle11\rangle$
coupling mode at a relatively low misorientation, i.e., at a misorientation
that is large enough for $\langle10\rangle$ dislocation cores to
overlap but smaller than $45^{\circ}$. By contrast, for the lower
$\epsilon$ value, the PN barriers are much smaller and of comparable
magnitude for the $\langle10\rangle$ and $1/2\langle11\rangle$ dislocations.
In this case, one would expect the transition between the two coupling
modes to occur at a misorientation angle close to $45^{\circ}$. Our
PFC simulation results reported below will confirm these expectations
by showing that the switching angle between the coupling modes is
close to $20^{\circ}$ when the PN barrier is much larger for $\langle10\rangle$
than for $1/2\langle11\rangle$ dislocations ($\epsilon=0.25$) and
comes close to $45^{\circ}$ when both barriers are negligibly small
($\epsilon=0.05$).

We note that the calculation of the PN barriers in PFC simulations
requires choosing the shearing velocity $v$ sufficiently small for
the viscous-like stress associated with grain translation (absent
in the MD simulations and real solids) to be much smaller than the
Peierls stress. Computation of this stress and reformulation of the
PFC model to eliminate this spurious viscous stress will be discussed
elsewhere \citep{Adland-2012b}. What is important for the present
study is that $v$ was chosen small enough for the results not to
depend on this PFC artifact.

The dynamical equations were solved using a pseudo-spectral scheme
with a mesh spacing $\Delta x=\Delta y=2\pi/8$ and a time step $\Delta t=0.1$.
This scheme extends the one described in \citep{Mellenthin08} to
treat the second order partial derivative in time in Eq.~(\ref{PFCdyn}).
Details of this scheme will be described elsewhere \citep{Adland-2012}.
Since the edge of the square unit cell with one atom per cell is $a=2\pi$
in dimensionless PFC units (and $\sqrt{2}a$ for the expanded unit
cell with 2 atoms used for comparisons with MD), this choice of mesh
spacing corresponds to having $8\times8=64$ mesh points per unit
cell of the square lattice. For the parameters used to melt and pull
the top and bottom surfaces of the crystal, we chose $b=5a=10\pi$,
$\xi=2a$, $\sigma=a=2\pi$, and $d=12a$.

\section{Equilibrium grain boundary structures}

Examples of equilibrium GB structures studied in this work are given
in Figures \ref{fig:GB-Structure-MD} and \ref{fig:Sigma-5}. Fig.~\ref{fig:GB-Structure-MD}(a)
shows an asymmetrical boundary with relatively small angles of $\theta=16.26^{\circ}$
and $\phi=14.04^{\circ}$. In the structure obtained by the MD simulations,
individual GB dislocations can be easily distinguished and their Burgers
vectors can be readily determined by the standard Burgers circuit
construction. This boundary has a periodic structure comprising six
$\langle100\rangle$ dislocations and four $1/2\langle110\rangle$
dislocations in each period. The core of each dislocation can be considered
as a stack of identical structural units representing capped trigonal
prisms in three dimensions. Their 2D projections appear as kites and
are outlined in the figure in green color. The PFC simulations give
a similar structure of this boundary, with the same number of dislocations
of each type (Fig.~\ref{fig:GB-Structure-MD}(b)). The exact positions
of the dislocations are different and vary with temperature and time.
However, the dislocation content predicted by both methods is identical.
In fact, all GBs with $\theta=16.26^{\circ}$ studied in this work
were found to be mixtures of these two types of dislocations in proportions
dictated by the inclination angle. These proportions were found to
be precisely the same in both MD and PFC simulations. In the particular
case of $\phi=0$, the boundary becomes symmetrical and its structure
represents an array of $\langle100\rangle$ dislocations. Likewise,
both methods confirm that the symmetrical boundaries with $\phi=\pm45{}^{\circ}$
are composed of $1/2\langle110\rangle$ dislocations.

As examples of high-angle GBs, Figs.~\ref{fig:Sigma-5}(a,b) show
the structures of symmetrical boundaries with $\theta=36.87^{\circ}$.
This lattice misorientation produces a CSL known as $\Sigma5$, $\Sigma$
being the reciprocal density of coincident sites \citep{Balluffi95}.
The structures shown in this figure are for the inclination angles
$\phi=0$ and $\phi=45{}^{\circ}$, corresponding to the GB planes
(310) and (210), respectively. Both structures are formed by topologically
identical kite-shape structural units similar to those in Fig.~\ref{fig:GB-Structure-MD}(a).
Such units are separated by one atomic bond when $\phi=0$ and are
connected head-to-tail when $\phi=45{}^{\circ}$. The rows of such
structural units running parallel to the tilt axis are similar to
dislocation cores in low-angle boundaries and can be also considered
as GB dislocations in high-angle boundaries.

Figure \ref{fig:Sigma-5}(c) illustrates a typical structure of an
asymmetrical $\Sigma5$ GB, with the inclination angle of $\phi=14.04^{\circ}$.
The kite-shape structural units ``clustered'' together or separated
by a bond can be considered as patches (facets) of the symmetrical
$\Sigma5\:(210)$ ($\phi=45{}^{\circ}$) and $\Sigma5\:(310)$ ($\phi=0$)
boundaries, respectively.

The PFC simulations show similar structural trends of high-angle GBs.
However, a detailed one-to-one comparison between the PFC and MD structures
is limited because of their different dimensionality. In particular,
the characteristic kite-shape structural units forming the 3D GBs
include sites located in adjacent (002) planes. Such structural units
have no analog in the 2D PFC structures containing only one layer.

\section{Mechanisms of grain boundary motion }

\subsection{MD results}

The mechanisms of GB motion will be discussed for relatively low-angle
boundaries in which the evolution of individual dislocations could
be reliably traced. Examination of MD snapshots revealed that the
GB motion at low temperatures ($0.3T_{m}$ to $0.5T_{m}$) was accomplished
by dislocation glide assisted by dislocation rearrangements and reactions.
As mentioned in Section \ref{sec:Introduction}, stress-driven motion
of symmetrical tilt GBs occurs by collective glide of identical dislocations
along parallel slip planes \citep{Cahn06a,Cahn06b}. This motion does
not require dislocation rearrangements or reactions. This mechanism
was indeed observed in our simulations as illustrated in Fig.~\ref{fig:Microrotation-vector-grain}(a)
for the boundary with $\theta=16.26^{\circ}$ and $\phi=0$.

In the case of asymmetrical GBs, the two types of dislocations forming
the GB structure belong to different slip systems. After a period
of time, the dislocations gliding in intersecting slip planes can
create locked configurations preventing further GB motion. Nevertheless,
our MD simulations have shown that the dislocations usually find a
way to glide past each other without completely locking themselves.
Fig.~\ref{fig:Microrotation-vector-grain}(b) illustrates the simultaneous
glide of $\langle100\rangle$ and $1/2\langle110\rangle$ dislocation
arrays during the motion of an asymmetrical boundary with $\theta=16.26^{\circ}$
and $\phi=38.66^{\circ}$. Note how the dislocation slip traces cross
each other multiple times without locking.

Two mechanisms were identified by which the dislocations could avoid
blocking each other while preserving the total Burgers vector. These
mechanisms involve dislocation reactions and dislocation avoidance,
respectively.

The dislocation reaction mechanism is similar to the one observed
in the recent MD study of shrinkage and rotation of isolated cylindrical
grains \citep{Trautt-2012}. Fig.~\ref{fig:Dislocation-reactions-a}
schematically illustrates a typical dislocation reaction process in
which a single $1/2\left\langle 110\right\rangle $ dislocation propagates
through an array of $\left\langle 100\right\rangle $ dislocations.
The structure shown in this figure represents a typical asymmetrical
GB with relatively small angles $\theta$ and $\phi<\theta$. At each
step of this process, the $\left\langle 100\right\rangle $ dislocation
on the immediate right of the $1/2\left\langle 110\right\rangle $
dislocation dissociates in two $1/2\left\langle 110\right\rangle $'s.
One of the product dislocations glides over a short distance (comparable
to the dislocation spacing) and recombines with the initial $1/2\left\langle 110\right\rangle $
dislocation to form a new $\left\langle 100\right\rangle $. The remaining
product dislocation is similar to the initial $1/2\left\langle 110\right\rangle $
but is located a short distance to its right (compare configurations
(a) and (g) in Fig.~\ref{fig:Dislocation-reactions-a}). This new
configuration continues to propagate to the right by repeating the
same steps: reaction of the $1/2\left\langle 110\right\rangle $ dislocation
with a neighboring $\left\langle 100\right\rangle $ and recombination
with one of its products. This process looks as if the $1/2\left\langle 110\right\rangle $
dislocation migrated along the boundary to the right (compare the
initial (a) and final (h) configurations in Fig.~\ref{fig:Dislocation-reactions-a}).
However, it is only the $1/2\left\langle 110\right\rangle $ Burgers
vector that propagates over many steps, whereas each individual dislocation
glides only over a short distance. Note that this dislocation propagation
process produces a slight displacement of the dislocation array down.
Multiple dislocation passes can produce significant GB displacements.
The remarkable feature of this mechanism is that it does not require
dislocation climb, despite the fact that the propagating dislocation
has a Burgers vector component normal to the GB plane.

A similar chain or reactions can propagate a $\left\langle 100\right\rangle $
dislocation through an array of $1/2\left\langle 110\right\rangle $'s
or $\left\langle 100\right\rangle $'s, or a $1/2\left\langle 110\right\rangle $
dislocation through an array of $1/2\left\langle 110\right\rangle $'s.
Such dislocation reactions provide a mechanism for rapid redistribution
of dislocation content over the GB without altering the total dislocation
content or relying on diffusion-controlled mechanisms such as climb.
These dislocation reactions prevent the formation of locks and simultaneously
provide a mechanism for GB motion.

The second mechanism was the dislocation avoidance. When the ratio
of the numbers of the two type of dislocations was large, we observed
that the minority dislocations tended to lag behind the majority dislocations
and then return to the boundary when a suitable gap was available.
This process is illustrated schematically in Fig.~\ref{fig:disloc-dissociation}
for the dislocation ratio 2:1. First, the majority dislocations move
forward while the minority dislocations are left behind. This separation
of dislocations creates gaps in the array of majority dislocations.
(Such gaps represent elastically distorted perfect lattice regions
between terminations of dislocation arrays and can be considered as
disclination dipoles \citep{Kleman2008}). Then, when the slip planes
of the minority dislocations come to alignment with the gaps, they
quickly glide forward and fill the gaps, recreating the initial GB
structure. As evident from the geometry of this process (Fig.~\ref{fig:disloc-dissociation}),
each minority dislocation fills a gap left by a neighboring minority
dislocation. Thus, the net result of this process is similar to the
dislocation reaction mechanism: both processes lead to relative translations
of two dislocation arrays parallel to the GB plane without interfering
with each other. Fig.~\ref{fig:MD-avoidance} illustrates the dislocation
avoidance mechanism for a particular GB with $\theta=16.26^{\circ}$
and $\phi=2.73^{\circ}$. This Figure shows that the minority dislocations
do not necessarily fill all gaps simultaneously but can fill them
one or several at a time.

The chains of dislocation reactions represent the dominant mechanism
responsible for the motion of asymmetrical GBs. The dislocation avoidance
mechanism was observed only for GBs with small inclinations and inclinations
close to $\pm45^{\circ}$, which contained large ratios of the different
dislocation types (e.g., 10:1). In a small number of simulations at
temperatures below $0.3T_{m}$, the dislocations formed strongly locked
configurations. Such locks eventually triggered generation of new
dislocations which initiated plastic deformation of the grains.

Similar mechanisms were found to operate in high-angle GBs, with rows
of kite-shape structural units playing the role of dislocation cores.

\subsection{PFC results\label{sub:PFC-mechanisms}}

In the PFC simulations, the dislocations forming the GBs with relatively
low angles could be readily identified and followed during the GB
motion. As expected, symmetrical boundaries with $\phi=0$ and $\phi=\pm45{}^{\circ}$
migrated by glide of identical dislocations in their respective slip
planes. Fig.~\ref{fig:PFC-disloc-trace1} shows a typical dislocation
trace during the motion of a GB with $\theta=16.26^{\circ}$ and $\phi=45{}^{\circ}$.
This boundary is composed of $1/2\left\langle 11\right\rangle $ dislocations
which glide in $(11)$ planes. A more detailed picture of this process
is illustrated in Fig.~\ref{fig:PFC-disloc-glide-(11)}. The dislocation
moves by a conservative process in which the structural units forming
its core are continually distorted and converted to perfect-lattice
units left behind the dislocation.

For asymmetrical GBs, however, the migration mechanisms are more complex.
Typically, the majority dislocations glide in their respective slip
planes as before, whereas the minority dislocations progress parallel
to the \emph{same} slip planes as the majority. This process is illustrated
in Fig.~\ref{fig:PFC-disloc-trace2} by dislocation traces in the
asymmetrical GB with $\theta=16.26^{\circ}$ and $\phi=30.2{}^{\circ}$.
This particular boundary contains one $\left\langle 10\right\rangle $
dislocation per every three $1/2\left\langle 11\right\rangle $ dislocations.
The majority dislocations move by perfect glide along $(11)$ planes
containing their Burgers vector. The minority $\left\langle 10\right\rangle $
dislocations move parallel to the same $(11)$ planes, a process which
cannot be explained by perfect glide. Indeed, the dislocation Burgers
vector has a component normal to $(11)$ planes, suggesting that the
dislocation motion must be accompanied by some amount of climb.

The latter conclusion was verified by observations of the detailed
core structure of the minority dislocations during their motion. As
illustrated in Fig.~\ref{fig:PFC-climb}, the number of sites (atomic
density peaks) in the dislocation core is not conserved. In this particular
example the core loses one site, whereas in other cases it could gain
sites. As noted in Section \ref{sec:Methodology-PFC}, PFC simulations
model an open system, in which the sites are not conserved locally
or globally. The PFC simulations represent the material's behavior
at high temperatures approaching the melting point. Therefore, the
continual creation, disappearance and redistribution of the sites
can be interpreted as occurring by diffusion of vacancies. This interpretation
can explain the motion of the minority dislocations along the majority
slip planes by a combination of glide and climb. It should be emphasized
that this mechanism was not, and could not be observed by the MD simulations
because of the short time scale.

As mentioned earlier, the 2D character of the PFC methodology precludes
a direction comparison of the structural units in high-angle GBs with
their MD counterparts. Furthermore, the non-conservative nature of
the PFC simulations often obscures unambiguous interpretation of atomic
movements in complex GB structures. These complications prevented
us from a more detailed PFC study of structural evolution and migration
mechanisms of asymmetrical high-angle GBs.

\section{Orientation and temperature dependencies of the driving stress}

Previous MD simulations \citep{Mishin07a,Ivanov08a} have shown that
the resistance of symmetrical tilt GBs to coupled motion is relatively
small and is due primarily to the stick-slip friction associated with
nucleation of disconnection loops. The present MD simulations indicate
that for asymmetric GBs, the resistance to motion is much greater
and is caused by the need to avoid or overcome locked configurations
between the dislocations gliding in intersecting slip planes. This
resistance can be characterized by the average shear stress $\sigma_{xy}$
required for sustaining a constant GB velocity.

Typical time dependencies of the shear stress are shown in Fig.~\ref{fig:stress-time}
for an asymmetrical GB in Al. The initial rise of the stress is due
to accumulation of elastic deformation until the stress reaches a
level sufficient for sustaining the GB motion. The values of $\sigma_{xy}$
reported below were obtained by averaging over the steady state portion
of the simulation run. At high temperatures the stress behavior is
uniformly noisy, whereas low temperatures reveal multiple peaks characteristic
of lock-unlock dynamics. It should be mentioned that in the case of
perfectly regular stick-slip behavior, the average stress depends
on the system size $L_{Y}$ in the direction normal the GB plane \citep{Ivanov08a}.
However, the size-dependent correction to the stress decreases as
$1/L_{Y}$ and for the large system sizes studied here is small.

The angle dependence of the average shear stress is plotted in Fig.~\ref{fig:stress-angle}
for a series of GBs with the same tilt angle $\theta=16.26^{\circ}$
but varying inclination angle. Observe that the stress is highest
for the ``most asymmetrical'' GBs with the inclination angles $\left|\phi\right|\approx20^{\circ}\textnormal{-}30^{\circ}$.
By contrast, the stresses required for moving the perfectly symmetrical
GBs arising at $\phi=0$ and $\pm45^{\circ}$ are almost an order
of magnitude smaller.

Fig.~\ref{fig:stress-angle} demonstrates that the driving stress
strongly decreases with temperature while preserving the same trend
of the angle dependence. A more detailed temperature dependence of
the stress is illustrated in Fig.~\ref{fig:Coupling-vs-Temperature}
for an asymmetrical GB in Al. The rapid decrease of the stress with
temperature reflects the thermally activated nature of the mechanisms
responsible for the dislocation reactions, avoidance and unblocking.
At high temperatures approaching the melting point, the driving stress
is a factor of ten smaller than at room temperature.

\section{Orientation and temperature dependencies of the coupling factor}

For symmetrical GBs, the coupling factor depends on the tilt angle
$\theta$ and temperature. As discussed in Section \ref{sec:Multiplicity},
previous MD simulations and experiments have revealed that $\beta$
is a multi-valued function of $\theta$ and exhibits a discontinuous
transition between two coupling modes. This transition was also confirmed
in the present MD simulations (not shown here). The existence of two
coupling modes and a discontinuity were also reproduced by the PFC
simulations as illustrated in Fig.~\ref{fig:beta-theta}. The jump
of the coupling factor occurs at an angle close to $20^{\circ}$ for
$\epsilon=0.25$, which is the value of $\epsilon$ used in all the
simulations of asymmetrical GBs, and close to $45^{\circ}$ for $\epsilon=0.05$.
This difference reflects the $\epsilon$-dependence of the PN barriers
of the two types of dislocations as discussed earlier. Note the excellent
agreement with predictions of the geometrical model of coupling \citep{Cahn06b},
in which the two branches of the plot are described by Eqs.~(\ref{eq:mode1})
and (\ref{eq:mode2}).

For asymmetrical GBs, we first discuss the $\theta=16.26^{\circ}$
GBs which were studied in greatest detail. Fig.~\ref{fig:beta-Cu-16}
reports the MD results for such boundaries in Cu and Al, showing two
temperatures in each case. As indicated in Section \ref{sec:Multiplicity},
for $\theta$ below the critical angle of approximately $36^{\circ}$,
$\beta$ is expected to remain a positive constant equal to the ideal
value given by Eq.~(\ref{eq:mode1}). Instead, the simulations show
that $\beta$ of asymmetrical GBs varies with the inclination angle
and, to a lesser degree, with temperature. The deviations from the
ideal value of $\beta$ are positive for some inclinations and negative
for others. In Al the deviations can be as high as a factor of two.
Al is dominated by positive deviations whereas in Cu both positive
and negative deviations occur to nearly equal extent.

Similar results were obtained by PFC simulations (Fig.~\ref{fig:PFC-beta-16}).
In this case, the angle of discontinuity of symmetrical GBs is about
$20^{\circ}$ (Fig.~\ref{fig:beta-theta}), thus for $\theta=16.26^{\circ}$
the coupling factor was again expected to remain constant. Instead,
it varies with $\phi$ in a manner reminiscent of that in Al (cf.
Fig.~\ref{fig:beta-Cu-16}(b)). The deviations from the ideal $\beta$
are always positive and reach a peak at about $\pm30^{\circ}$. It
should be emphasized that, despite the significant deviations of $\beta$
from its ideal value in both MD and PFC simulations, the coupling
factor remains positive. The positive sign of $\beta$ indicates that
the coupling mode remains the same at all inclination angles, which
is consistent with our geometric analysis in Section \ref{sec:Multiplicity}.

Fig.~\ref{fig:beta-Cu-37}(a) shows the MD results for high-angle
GBs with $\theta=36.87^{\circ}$. In this case, the coupling factor
is expected to be negative at $\phi=0$ and change sign to positive
values as $|\phi|$ increases towards $45^{\circ}$ (Section \ref{sec:Multiplicity}).
This behavior is indeed confirmed by the simulations, despite the
significant scatter of the points and deviations from the ideal values
of $\beta$ (Fig.~\ref{fig:beta-Cu-37}(a)). The scatter strongly
increases in the discontinuity regions, in which the magnitude of
$\beta$ becomes very large. Importantly, the PFC simulations reveal
a very similar behavior of the coupling factor (Fig.~\ref{fig:beta-Cu-37}(b)).
The PFC plot is smoother than the MD plot and clearly reveals a ``divergence''
of $\beta$ ($\beta\rightarrow\pm\infty$) in the narrow region where
the sign changes.

Due to the high computational efficiency of the PFC method, it was
possible to perform a large set of simulations with various misorientation
and inclination angles. The results are summarized in Fig.~\ref{fig:Summary-of-PFC}
as a diagram in the coordinates $\theta$-$\phi$ showing positive
and negative values of $\beta$ by different symbols. The boundary
between the positive and negative regions has been drawn in this figure
by hand and is intended to be a trend line. Despite the obvious scatter
of the points, the shape of this line is in qualitative agreement
with the geometric predictions shown in Fig.~\ref{fig:Schematic-modes}
as far as the overall shape of the two regions of positive and negative
$\beta$ is concerned. In particular, this diagram clearly shows that
the discontinuity between the two coupling modes known from the previous
work \citep{Suzuki05b,Cahn06a,Cahn06b,Molodov07b,Molodov09a,Gorkaya2009,Gorkaya2010,Molodov2011}
exists not only for symmetrical boundaries but also extends over the
full range of inclination angles. However, sections of this diagram
at a constant misorientation angle in the MD (Figs.~\ref{fig:beta-Cu-16}
and \ref{fig:beta-Cu-37}(a)) and PFC (Figs.~\ref{fig:PFC-beta-16}
and \ref{fig:beta-Cu-37}(b)) simulations clearly show that the magnitude
of $\beta$ is strongly dependent on inclination in a way that cannot
be predicted by the geometrical considerations which were used to
construct Fig.~\ref{fig:Schematic-modes}.

\section{Discussion}

The goal of this paper was to investigate the effect of the inclination
angle on stress-driven motion of asymmetrical tilt GBs. To ensure
generality of the results, the MD simulations were conducted over
the full range of inclination angles, a wide temperature range, and
in two different metals. Furthermore, the same GBs were also studied
by PFC simulations, a recently developed methodology which provides
access to atomic-level processes on diffusive time scales \citep{Elder2002,Elder2004,Berry2006,Stefanovic2006,Wu-2007,Elder2007,Berry08,Mellenthin08,Stefanovic2009,Wu2010a,Jaatinen-2010,Spatschek2010,Olmsted2011,Adland-2012}.

The MD and PFC methods are complementary to each other and their combination
offers an efficient approach to multiscale problems such as GB motion.
MD simulations are suitable for studying the effects of temperature,
strain rate and stress on GB dynamics \citep{Mishin07a,Ivanov08a}
while simultaneously providing detailed information about atomic mechanisms.
However, dislocation climb and other diffusion-controlled processes
are beyond the time scale of MD simulations (tens of nanoseconds).
The PFC approach is less quantitative and cannot be used for tracking
all details of atomic movements. However, the material is modeled
in the long-time regime in which atomic diffusion can readily occur.
This gives access to GB migration mechanisms (such as dislocation
climb) that would otherwise not be seen.

This study has shown that at a fixed misorientation angle $\theta$,
the coupling factor $\beta$ varies with the inclination angle $\phi$
(Figs.~\ref{fig:beta-Cu-16}, \ref{fig:PFC-beta-16}, \ref{fig:beta-Cu-37}).
This observation does not confirm the geometric prediction \citep{Cahn06b}
that $\beta$ is a function of $\theta$ only. But there are other
theoretical predictions that were tested and verified by this work.
The most important of them is that most GBs in materials are coupled
and, unless subject to imposed constraints, can be moved by applied
stresses \citep{Cahn06b}. Indeed, the overwhelming majority of asymmetrical
GBs tested here coupled to shear stresses and were moved by them.
This finding clearly demonstrates that the coupling effect is \emph{not}
an attribute of only symmetrical GBs which were predominantly studied
in previous work.

Furthermore, our simulations show that the two coupling modes found
previously in {[}001{]} symmetrical tilt GBs \citep{Suzuki05b,Cahn06a,Cahn06b,Molodov07b,Molodov09a,Gorkaya2009,Gorkaya2010,Molodov2011}
continue to exist in asymmetrical GBs. The discontinuous transition
between the modes also continues to exist for asymmetrical GBs as
illustrated in Figs.~\ref{fig:Schematic-modes} and \ref{fig:Summary-of-PFC}.
In the vicinity of the transition between the modes, the magnitude
of the coupling factor becomes very large if not divergent (Fig.~\ref{fig:beta-Cu-37}).
In other words, a GB caught between the two coupling modes responds
to applied shears by a processes which appears like sliding. Further
investigations are needed to determine whether this response is a
manifestation of a true sliding process or of the ``dual behavior''
with alternation between the two coupling modes as suggested on p.~4974
of \citep{Cahn06b}.

The long-standing mystery \citep{Read50a} related to asymmetrical
boundaries is how the dislocations gliding in intersecting slip planes
avoid blocking each other (Fig.~\ref{fig:Microrotation-vector-grain}).
Strong interactions between such dislocations can explain the high
stresses needed for the coupled motion of asymmetrical GBs (Fig.~\ref{fig:stress-angle}).
At the same temperature and GB velocity, the shear stress required
for steady-state motion of an asymmetrical GB can be an order of magnitude
greater than for symmetrical boundaries. Several mechanisms have been
found by which the dislocations alleviate the locks, most notably
the dislocation reactions (Fig.~\ref{fig:Dislocation-reactions-a})
and dislocation avoidance (Figs.~\ref{fig:disloc-dissociation} and
\ref{fig:MD-avoidance}). When diffusion is allowed, the locks can
also be overcome by dislocation climb as suggested by the PFC simulations
(Section \ref{sub:PFC-mechanisms}). The operation of these complex
mechanisms responsible for the unlocking of the dislocations can explain
the deviations of the coupling factor from the ideal geometrical value
(Figs.~\ref{fig:beta-Cu-16}, \ref{fig:PFC-beta-16} and \ref{fig:beta-Cu-37}).
Such deviations are predominantly positive, suggesting the existence
of a sliding component along with coupling. However, a better understanding
of the origin of the deviations requires further studies.

One might think that the deviations of the coupling factors from the
ideal values could be caused by an additional driving force arising
due to the elastic anisotropy of the lattice. For asymmetrical GBs,
the elastic anisotropy creates a difference in elastic strain energies
in the grains which may drive the GB motion. In fact, this driving
force underlines one of the methods for studying GB migration by atomistic
computer simulations, see e.g. \citep{Mishin2010a} for review. However,
because this driving force is quadratic in stress, large stresses
need to be created in the grains in order to drive GB motion over
a non-negligible distance. The stresses used in our study were smaller
and unlikely to explain the above deviations. This is also evident
from Fig.~\ref{fig:Coupling-vs-Temperature} showing the temperature
dependencies of the coupling factor and stress. At temperatures above
500 K, $\beta$ does not change within the scatter of the data points
and remains higher than the ideal value. In the same temperature interval,
the stress drops by about an order of magnitude, which rules out the
possibility that the deviation from the ideal $\beta$ was caused
by additional GB motion induced by the stress.

It is interesting to compare the predictions of this modeling study
with experimental data. Our main finding that most of asymmetrical
GBs are moved by applied shear stresses is well consistent with experimental
observations of stress-driven grain growth in nano-crystalline materials
\citep{Weertman05a,Hemker07a,Legros08,Gianola08a}. Coupled motion
of asymmetrical tilt GBs was also observed in experiments on bicrystalline
samples subject to applied shear loads \citep{Molodov2011,Syed2012}.
In particular, Molodov et al.~\citep{Molodov2011} studied stress-driven
motion of {[}001{]} tilt GBs in Al bicrystals with different crystallographic
parameters. One of the experiments was performed on a bicrystal with
$\theta=17.4{}^{\circ}$ and $\phi=19.1{}^{\circ}$. The boundary
was found to couple to the applied stress (see Figure 7 in \citep{Molodov2011})
and move with a coupling factor of 0.39, which is higher than the
ideal geometrical value 0.31 computed from Eq.~(\ref{eq:mode1}).
To our knowledge, this is the only published value of coupling factor
for an asymmetrical tilt GB in Al. This value can be \emph{directly}
compared with our simulations. Indeed, from the MD simulation results
for Al shown in Fig.~\ref{fig:beta-Cu-16}(b), the boundary closest
to the experimental conditions is one with $\theta=16.26{}^{\circ}$
and $\phi=18.44{}^{\circ}$. The computed coupling factor for this
boundary is 0.40 at the temperature of 500 K and 0.38 at 900 K. Although
the experimental temperature corresponding to $\beta=0.39$ was not
specified in \citep{Molodov2011}, our prediction is in excellent
agreement with experiment at both temperatures. Our simulations indicate
that the overestimated coupling factor is caused by the dislocation
reactions as discussed earlier in this paper. It should also be pointed
out that the overestimated (relative to the geometric prediction)
$\beta$ value found in \citep{Molodov2011} is well consistent with
the general trend found in this work by both MD (Fig.~\ref{fig:beta-Cu-16}(b))
and PFC (Fig.~\ref{fig:PFC-beta-16}) simulations.

The symmetry analysis (Sections \ref{sec:symmetry} and \ref{sec:Multiplicity})
and the simulations reported in this paper provide a theoretical basis
for designing future experiments. In particular, for {[}001{]} tilt
GBs in FCC metals, the following experiments would provide the most
direct test of our theoretical predictions. First, for a given material,
a series of symmetrical tilt GBs should be studied over the entire
range of misorientation to determine the critical switching angle
$\theta_{c}$ between the different coupling modes $\left\langle 100\right\rangle $
and $\left\langle 110\right\rangle $. This series of experiments
would parallel the experimental study of symmetric tilt GBs in Al
where $\theta_{c}$ was estimated to be between $30.5{}^{\circ}$
and $36.5{}^{\circ}$ \citep{Gorkaya2009}. After this critical angle
has been established, the following series of experiments should be
able to reveal the three distinct regimes predicted in this paper:
\begin{enumerate}
\item Measurements of $\beta$ at a fixed tilt angle $\theta<\theta_{c}$
and varying inclination angle $\phi$. The coupling factor is expected
to remain positive but may deviate from the value predicted by Eq.~(\ref{eq:mode1})
(most probably in the positive direction).
\item Measurements of $\beta$ at a fixed tilt angle $\theta>\left(90{}^{\circ}-\theta_{c}\right)$
and varying inclination angle $\phi$. The coupling factor is expected
to remain native but may deviate from the value predicted by Eq.~(\ref{eq:mode2})
(most probably in the negative direction).
\item Measurements of $\beta$ at a fixed tilt angle $\theta_{c}<\theta<\left(90{}^{\circ}-\theta_{c}\right)$
and varying inclination angle $\phi$. The coupling factor is expected
to switch from positive at small $\phi$ to negative as $\phi$ approaches
$\pm45^{\circ}$. At angles close to the sign change, the boundary
may exhibit a sliding-like behavior.
\end{enumerate}
It is important to emphasize that the critical switching angle between
the two coupling modes is essential in determining the range of misorientation,
$\theta_{c}<\theta<\left(90{}^{\circ}-\theta_{c}\right)$, which is
predicted to exhibit the most interesting and novel dependence of
the coupling on inclination. As indicated earlier, the specific critical
angle of $\theta_{c}=36{}^{\circ}$ mentioned above refer to EAM Cu
and can be different for other metals.

\section{Conclusions }

In conclusions, we have shown by computer simulations that the coupling
effect exists for the vast majority of symmetrical as well as asymmetrical
tilt GBs. Furthermore, as for symmetrical GBs, the coupling of asymmetrical
boundaries can occur in multiple models with coupling factors $\beta$
that can have different signs. The magnitude of $\beta$ is generally
strongly dependent on the inclination angle in a way that cannot be
fully predicted from purely geometrical considerations. The most dramatic
manifestation of this dependence is the sharp increase in the magnitude
of $\beta$ in a range of angles near the boundary between regions
of opposite signs of $\beta$. This confers GBs with a sliding-like
behavior that could potentially have a strong influence on mechanical
properties of polycrystalline materials.

Furthermore, we have found that the motion of asymmetrical GBs can
be mediated by several different processes. In the MD simulations
where vacancy diffusion was negligible, the motion of GB dislocations
on different slip planes was accommodated by dislocation reactions
and/or avoidance. In contrast, in the PFC simulations which incorporate
diffusive processes generally occurring at high temperatures, dislocation
climb facilitated collective motion of dislocations with different
Burgers vectors, allowing the GB to avoid locks and move smoothly. 

Importantly, the dependence of the coupling factor $\beta$ on GB
bicrystallography was found to be strikingly similar in the MD and
PFC simulations, despite the mentioned differences in the atomistic
details of GB migration. Both simulation methods predict similar shapes
of the regions of opposite signs of $\beta$ in the parameter space
of angles. These shapes are consistent with considerations of crystal
symmetry and the different PN barriers of the dislocations moving
in different slip planes. Both methods predict diverging magnitudes
of $\beta$ near the boundary separating the regions of positive and
negative values of $\beta$. We therefore expect these basic features
of asymmetrical GBs to pertain to a wide range of materials, temperatures
and other physical conditions.

Finally, our simulation results are in encouraging agreement with
experiments on stress-driven GB motion in polycrystalline materials
and bicrystalline samples. Furthermore, they are in excellent agreement
with the recently measured experimental value of the coupling factor
for an asymmetrical tilt GB in an Al bicrystal \citep{Molodov2011}.
When discussing the problem of coupled motion of asymmetrical boundaries,
Molodov et al.~\citep{Molodov2011} remark: ``Further investigations,
especially molecular dynamics simulations are obviously needed to
clarify the mechanisms of this phenomenon, specific atomic rearrangements,
dislocation processes, and reactions involved in the process of boundary
migration.'' We hope that the present study has made a step in this
direction and will motivate further experiments.

\emph{Acknowledgments -} This work was supported by the U.S. Department
of Energy, the Physical Behavior of Materials Program, through Grants
No. DE-FG02-01ER45871 (ZTT and YM) and DE-FG02-07ER46400 (AA and AK).

\bibliographystyle{unsrt}
\bibliography{cond-mat.mtrl-sci.bbl}

\newpage{}\clearpage{}

\begin{center}
\begin{figure}
\begin{centering}
\includegraphics[scale=0.5]{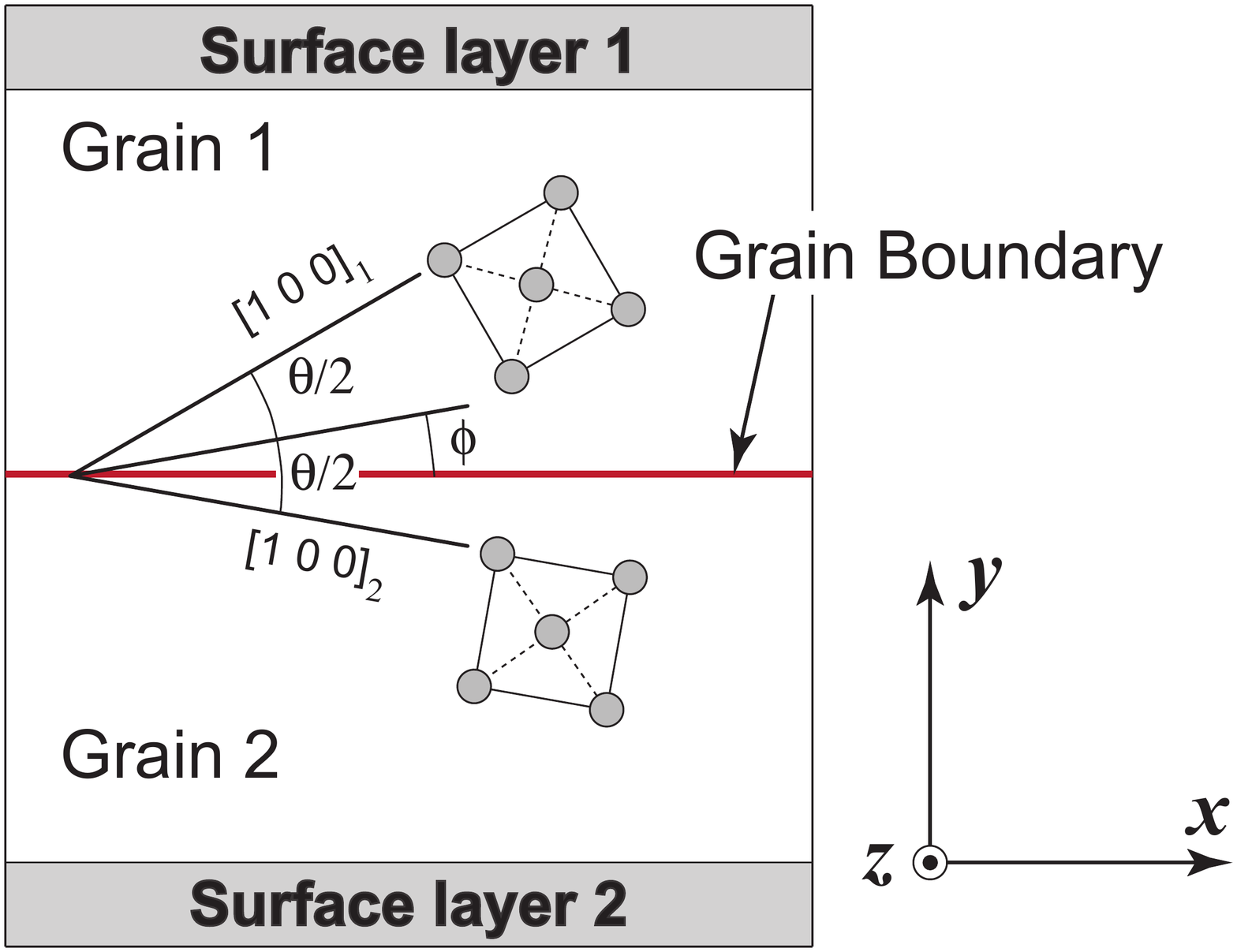} 
\par\end{centering}

\caption{Geometry of the MD simulation block employed in this study.\label{fig:Simulation-Geometry}}
\end{figure}

\par\end{center}

\begin{center}
\begin{figure}
\noindent \begin{centering}
\includegraphics[scale=0.9]{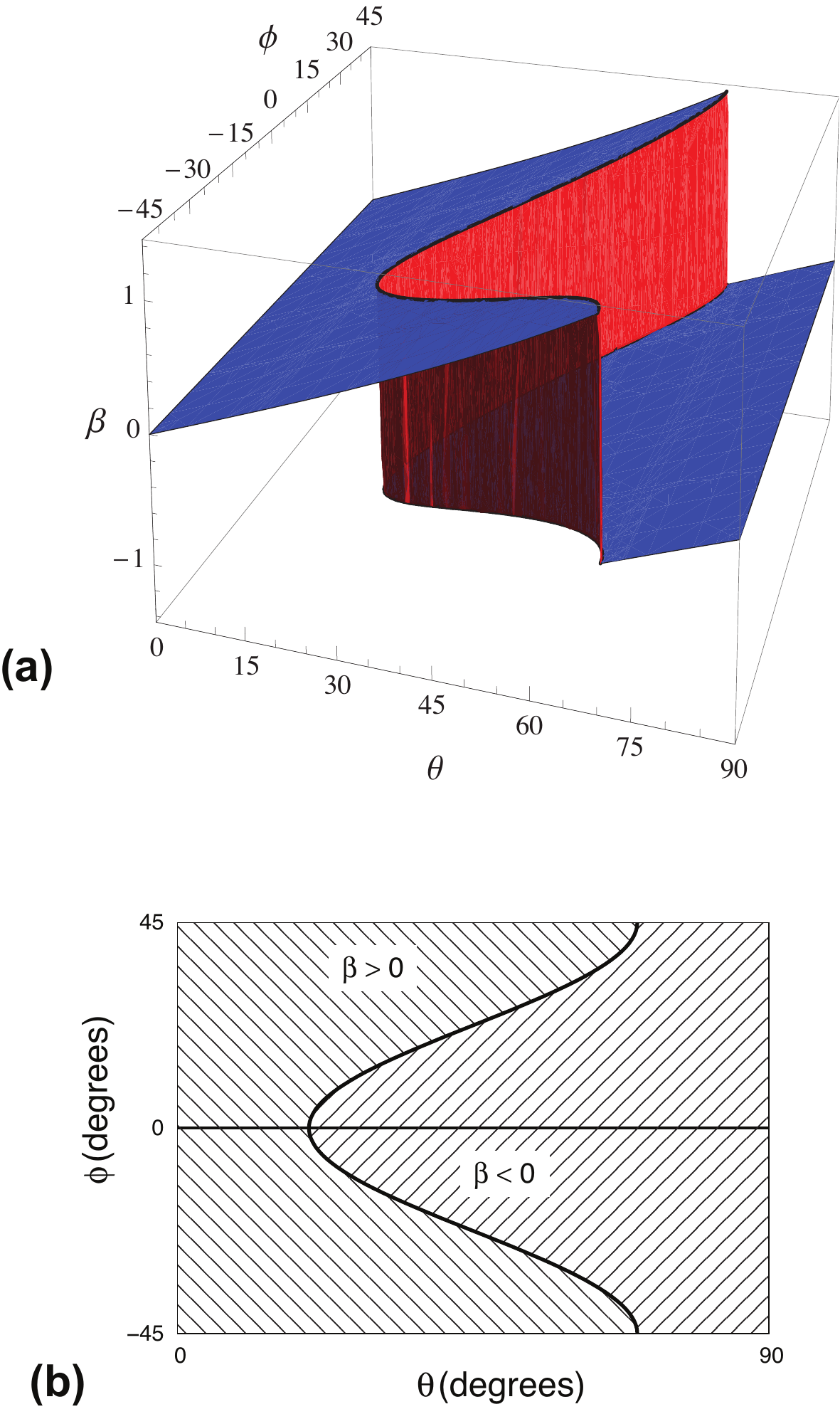} 
\par\end{centering}

\caption{Schematic dependence of the coupling factor $\beta$ on the tilt angle
$\theta$ and the inclination angle $\phi$. (a) Surface plot with
two sheets representing the $\left\langle 100\right\rangle $ and
$\left\langle 110\right\rangle $ modes of coupling. The coupling
factor is discontinuous along the cut. (b) Projection of the surfaces
on the $\theta$-$\phi$ plane, showing the areas of positive and
negative values of $\beta$.\label{fig:Schematic-modes}}
\end{figure}

\par\end{center}

\begin{center}
\begin{figure}
\begin{centering}
\includegraphics[clip,height=7in]{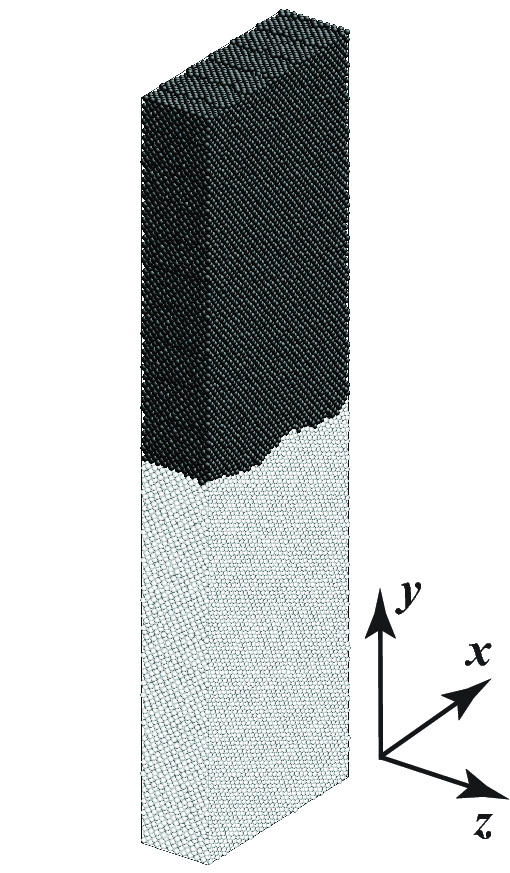} 
\par\end{centering}

\caption{Illustration of partitioning of atoms between the two grains using
the orientation parameter method. The image was taken from MD simulations
of the Cu GB with $\theta=16.26^{\circ}$ and $\phi=14.04^{\circ}$
at 500 K. The bright and dark colors designate atoms assigned to different
grains.\label{fig:Orientation-parameter}}
\end{figure}

\par\end{center}

\begin{center}
\begin{figure}
\noindent \begin{centering}
\includegraphics[scale=0.4]{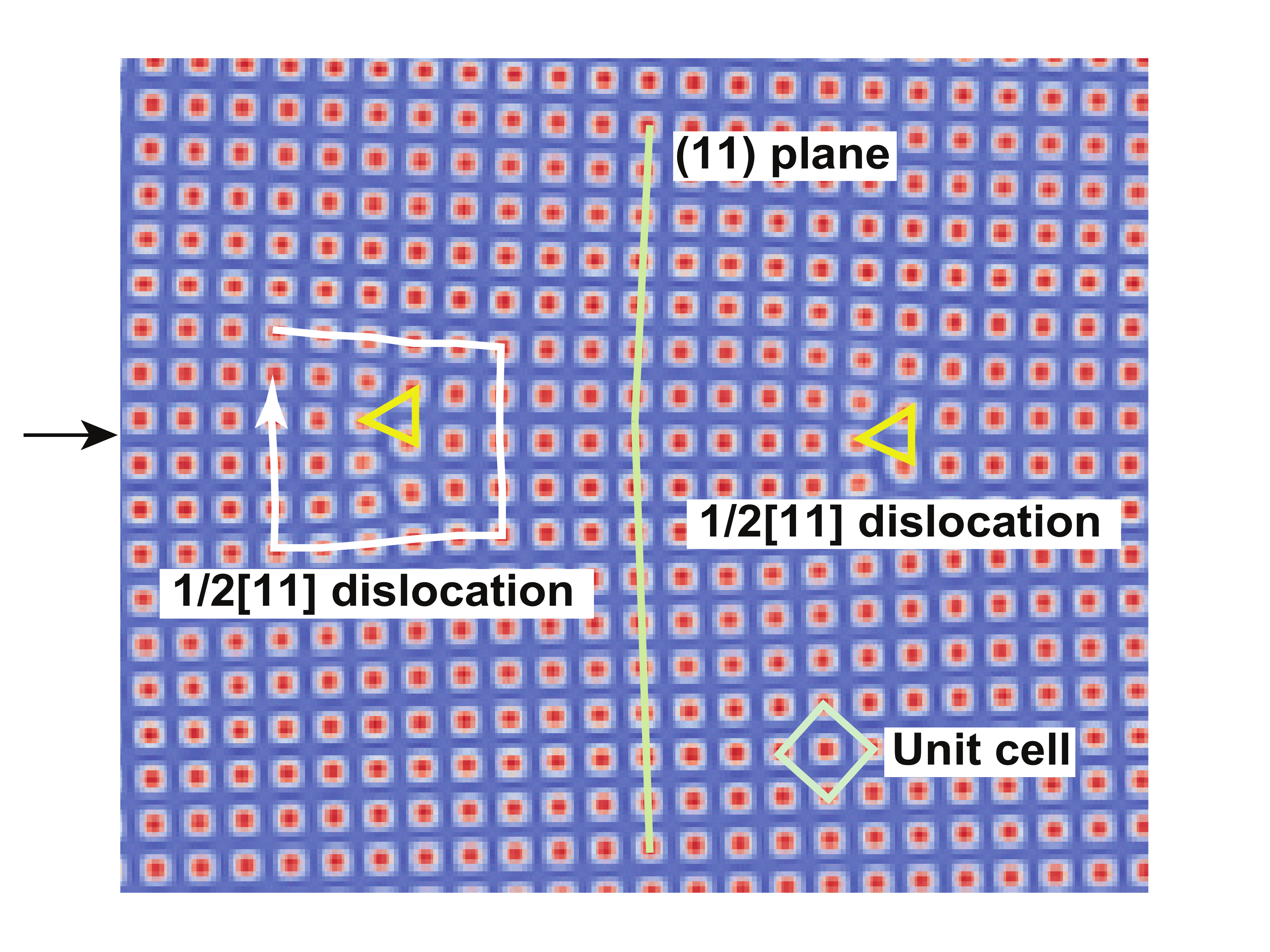} 
\par\end{centering}

\caption{PFC simulated structure of the symmetrical tilt GB with $\theta=80^{\circ}$.
The GB plane is horizontal and its approximate position is indicated
by an arrow. The structure is composed of $1/2\langle11\rangle$ dislocations
whose cores are marked by yellow triangles. A Burgers circuit drawn
around one of the dislocations is shown in white color. The unit cell
of the 2D lattice and a perfect (11) crystal plane passing between
the dislocations are indicated.\label{fig:PFC-80degrees} }
\end{figure}

\par\end{center}

\begin{center}
\begin{figure}
\begin{centering}
\includegraphics[scale=0.35]{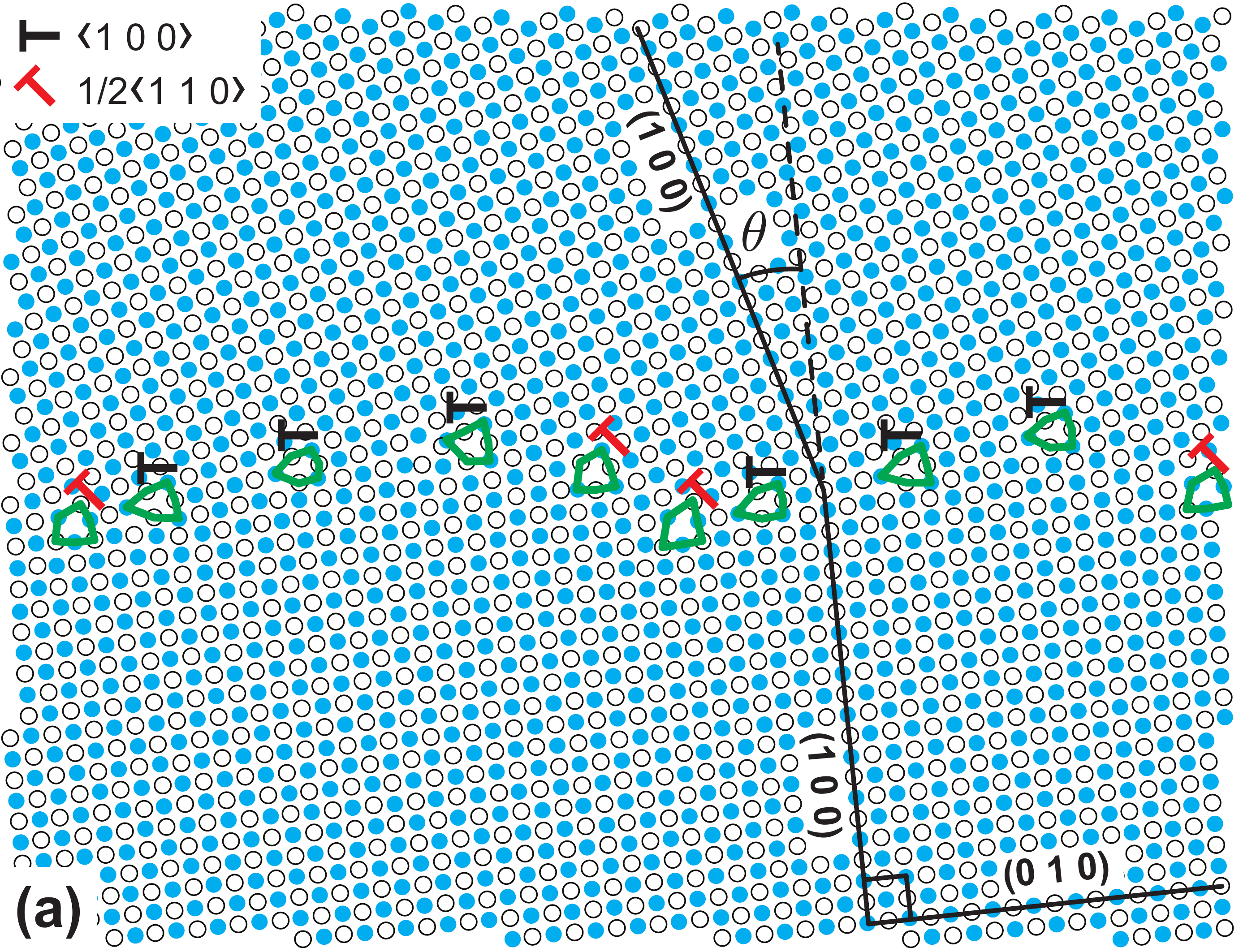} 
\par\end{centering}

\bigskip{}

\begin{centering}
\includegraphics[scale=0.35]{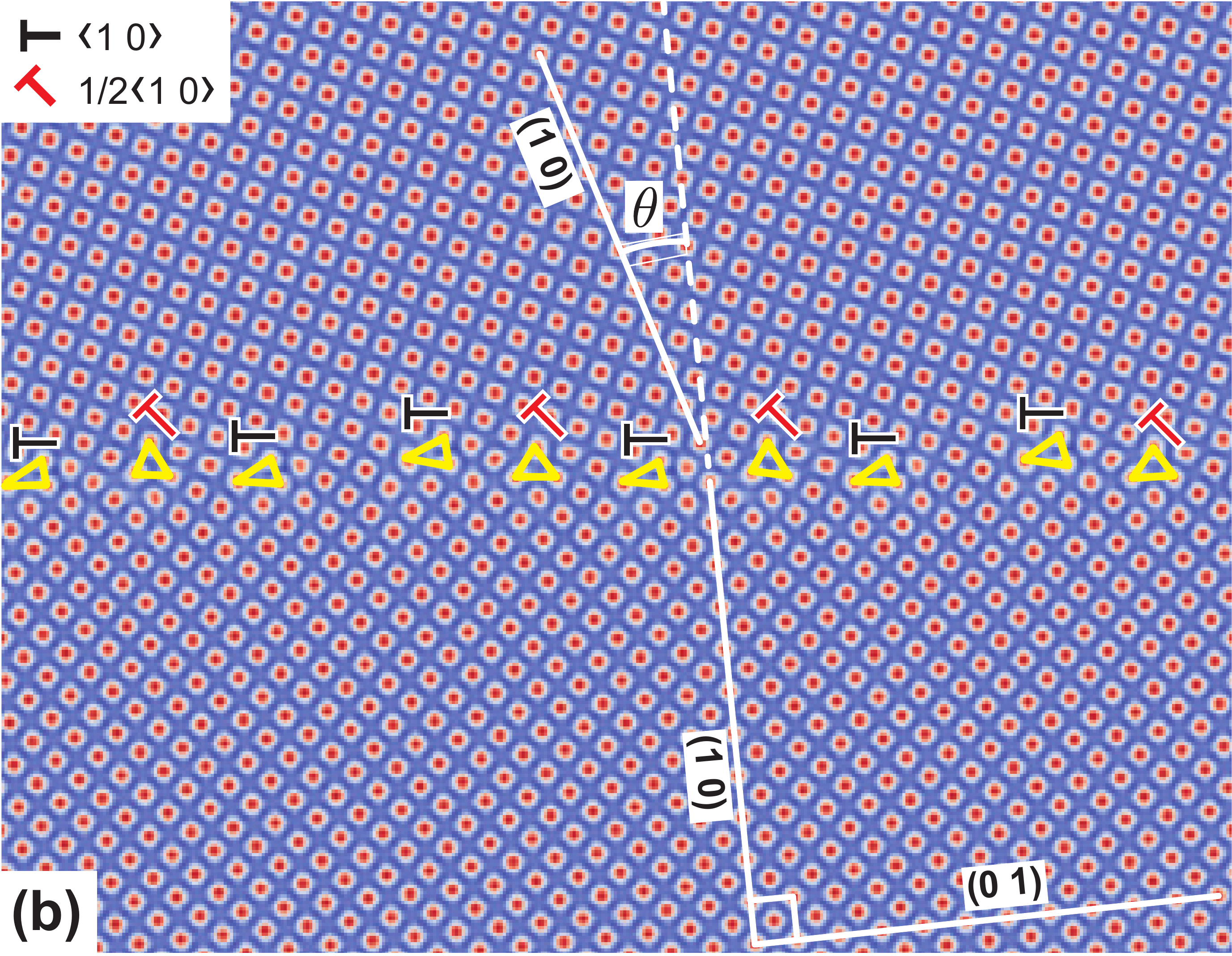} 
\par\end{centering}

\caption{Equilibrium structure of the asymmetrical GB with $\theta=16.26^{\circ}$
and $\phi=14.04^{\circ}$. (a) Obtained by MD simulations of Cu and
Al. The open and filled circles mark atomic positions in alternate
(002) planes. The structural units forming the dislocation cores are
outlined. (b) Obtained by PFC simulations. The dislocation cores are
marked by yellow triangles. Both structures are composed of six $\langle100\rangle$
dislocations and four $1/2\langle110\rangle$ dislocations in each
structural period. \label{fig:GB-Structure-MD}}
\end{figure}

\par\end{center}

\begin{center}
\begin{figure}
\noindent \begin{centering}
\includegraphics[scale=0.7]{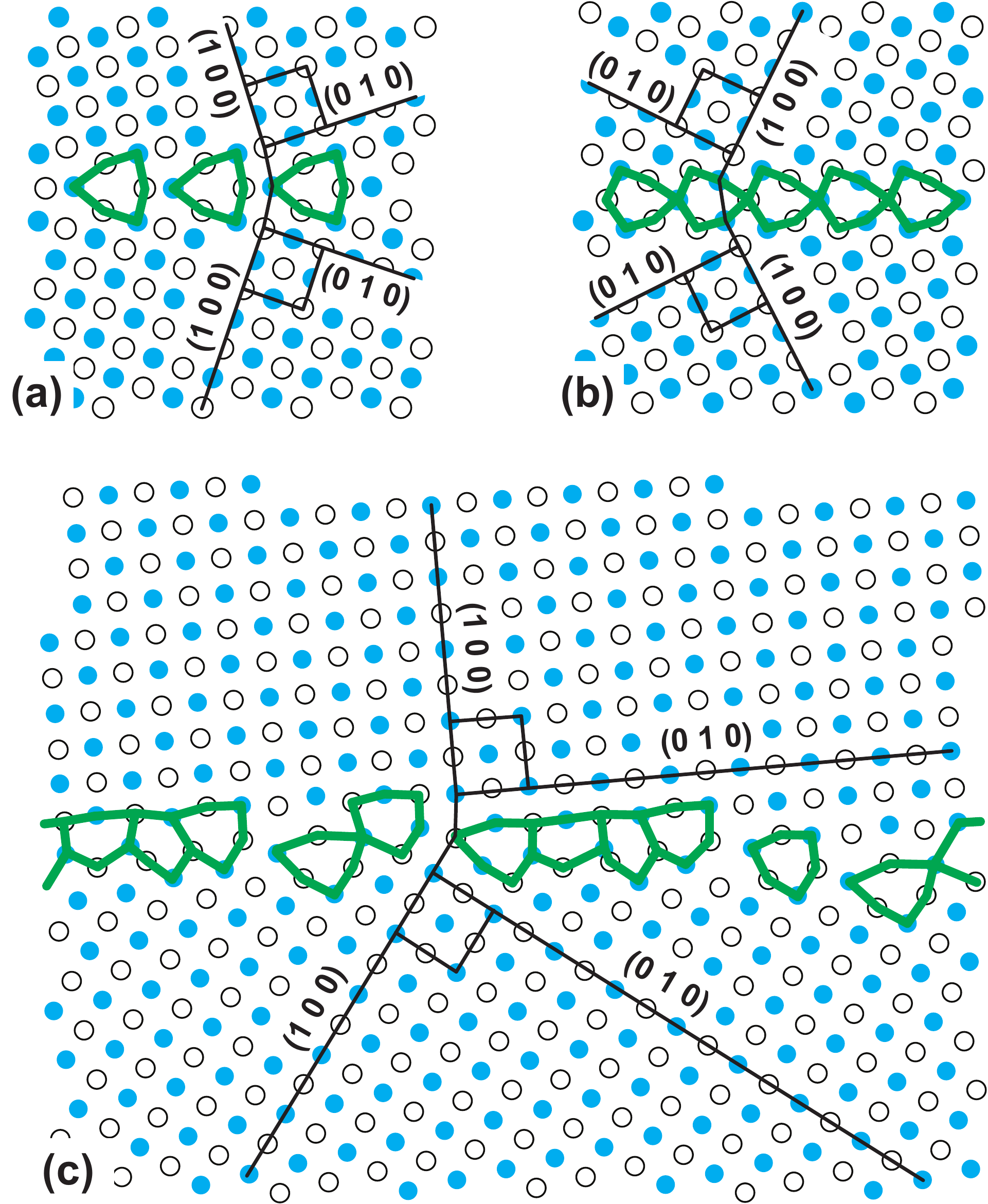} 
\par\end{centering}

\caption{Examples of equilibrium structures of $\Sigma5$ GBs ($\theta=36.87^{\circ}$)
obtained by MD simulations of Cu and Al. (a) Symmetrical (310) boundary
with $\phi=0$. (b) Symmetrical (210) boundary with $\phi=45{}^{\circ}$.
(c) Asymmetrical boundary with $\phi=14.04^{\circ}$. The kite-shape
structural units of the GB structures are outlined. Selected crystal
planes are labeled for clarity.\label{fig:Sigma-5}}
\end{figure}

\par\end{center}

\begin{figure}
\begin{centering}
\includegraphics[width=6in]{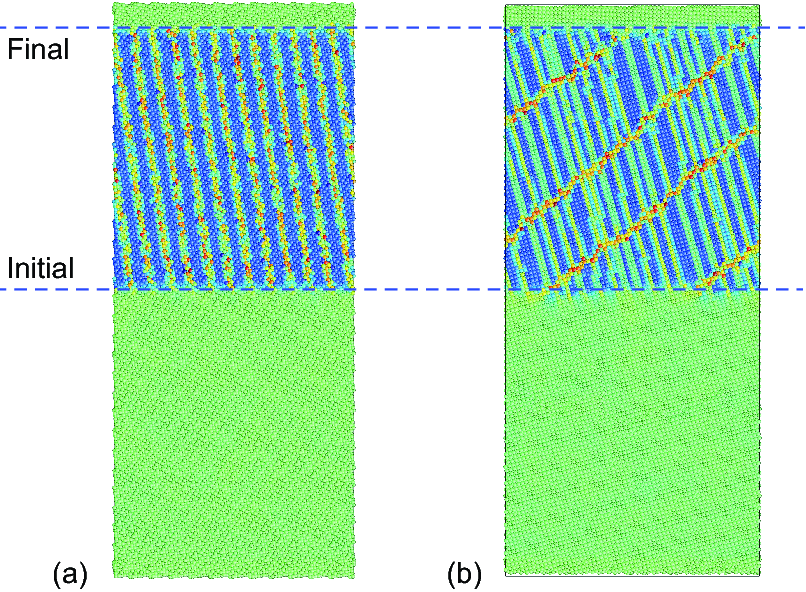} 
\par\end{centering}

\caption{Traces of gliding dislocations in MD simulations of coupled GB motion
in Cu at 500 K. The dislocation traces are revealed using the micro-rotation
vector method from \citep{Tucker2010}. The green and blue colors
represent different localized lattice deformations. The initial and
final GB positions are indicated. (a) Symmetrical tilt GB with $\theta=16.26^{\circ}$
and $\phi=0$ moves by collective glide of $\langle100\rangle$ dislocations.
(b) Asymmetrical tilt GB with $\theta=16.26^{\circ}$ and $\phi=38.06^{\circ}$
moves by collective glide and reactions of $\langle100\rangle$ (minority)
and $1/2\langle110\rangle$ (majority) dislocations. \label{fig:Microrotation-vector-grain}}
\end{figure}

\begin{figure}
\begin{centering}
\includegraphics[scale=0.5]{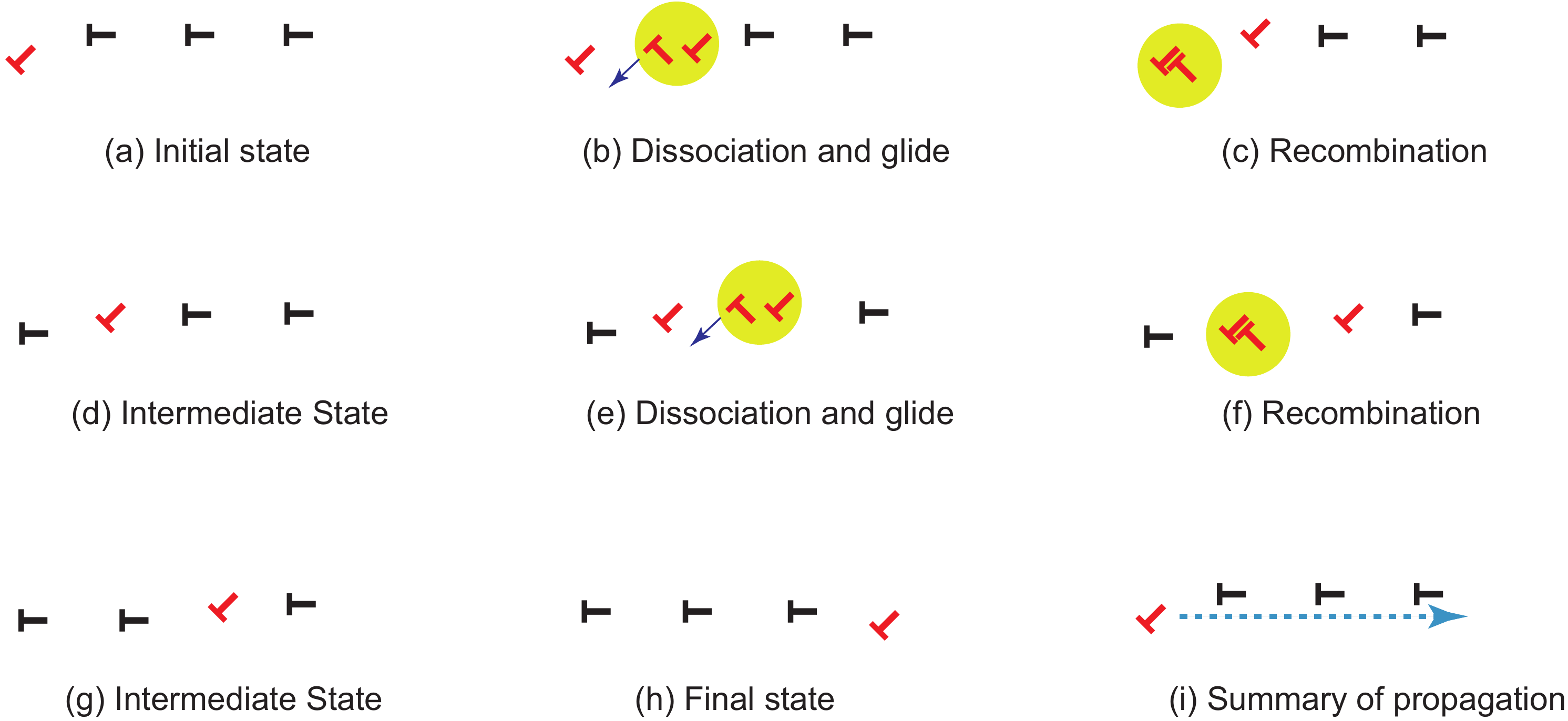} 
\par\end{centering}

\caption{Schematic of dislocation propagation along a GB by a chain of dislocation
reactions. The dislocation notation is the same as in Figure \ref{fig:GB-Structure-MD}.
The dislocation pairs undergoing dissociation and recombination reactions
are encircled. (a) Initial state. (d) and (g) Intermediate states.
(h) Final state. (i) Summary of the whole process. \label{fig:Dislocation-reactions-a}}
\end{figure}

\begin{figure}
\noindent \begin{centering}
\includegraphics[scale=0.8]{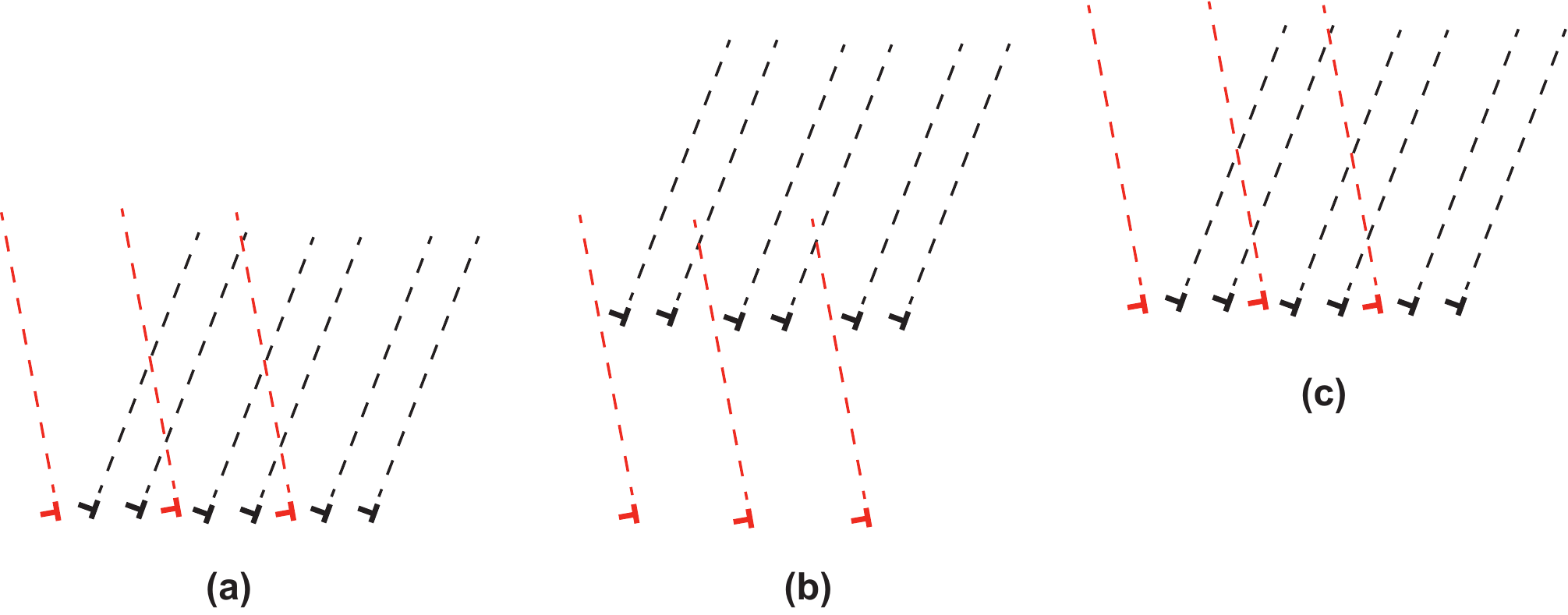} 
\par\end{centering}

\caption{Schematic illustration of the dislocation avoidance mechanism for
an asymmetrical GB moving upward. The dislocation notation is the
same as in Figure \ref{fig:GB-Structure-MD}. (a) Initial dislocation
structure comprising $\langle100\rangle$ (majority) and $1/2\langle110\rangle$
(minority) dislocations. The dashed lines indicate the dislocation
slip planes ahead of their motion. (b) The majority dislocations move
forward while the minority are left behind, creating gaps in the GB
structure. (c) When the gaps are aligned with the slip planes of the
minority dislocations, the latter move forward and fill the gaps,
recreating the initial GB structure in a new position.\label{fig:disloc-dissociation} }
\end{figure}

\begin{figure}
\noindent \centering{}\includegraphics[scale=1.41]{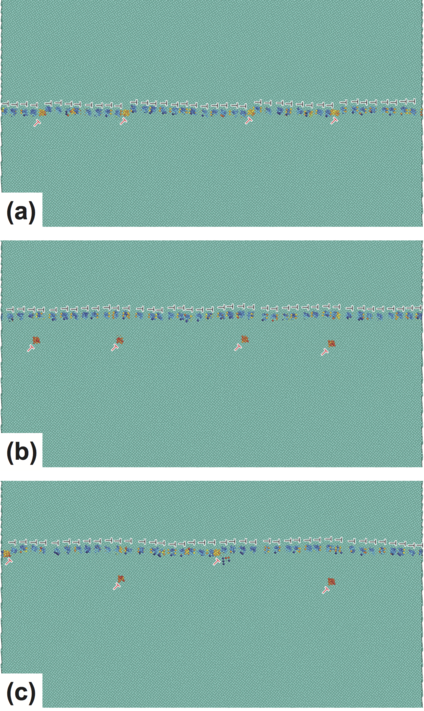}\caption{MD snapshots illustrating the dislocation avoidance mechanism for
an asymmetrical GB in Cu with $\theta=16.26^{\circ}$ and $\phi=2.73^{\circ}$.
The GB moves upward. (a) Initial GB structure. (b) The minority dislocations
$1/2\langle110\rangle$ are left behind creating gaps between the
majority dislocations $\langle100\rangle$. (c) When proper alignment
is reached, two of the minority dislocations catch up with the boundary
and fill the gaps. The remaining minority dislocations subsequently
fill other gaps (not shown). The dislocations are visualized by constructing
small Burgers loops around their cores and color-coding a selected
projection of the closure failure. The simulation temperature is 500
K. \label{fig:MD-avoidance}}
\end{figure}

\begin{figure}
\begin{centering}
\includegraphics[scale=0.8]{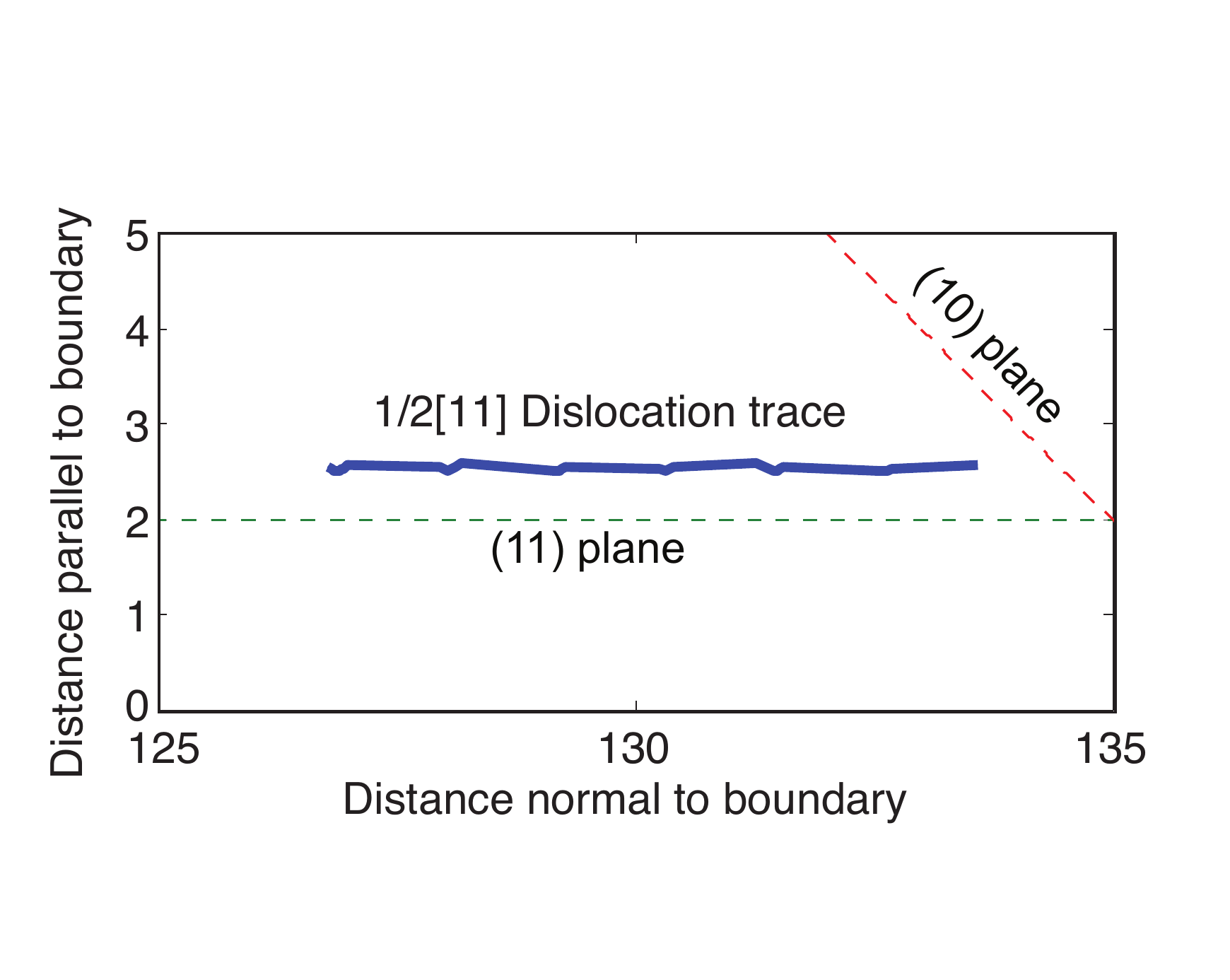} 
\par\end{centering}

\caption{A typical trace of a $1/2\langle11\rangle$ dislocation in PFC simulations
of coupled motion of the symmetrical GB with $\theta=16.26^{\circ}$
and $\phi=45{}^{\circ}$. The dashed lines indicate $(11)$ and $(10)$
crystal planes in the advancing grain. The dislocation moves parallel
to $(11)$ planes consistent with the glide mechanism. The distances
are in Angstroms.\label{fig:PFC-disloc-trace1}}
\end{figure}

\begin{figure}
\begin{centering}
\includegraphics[clip,scale=0.7]{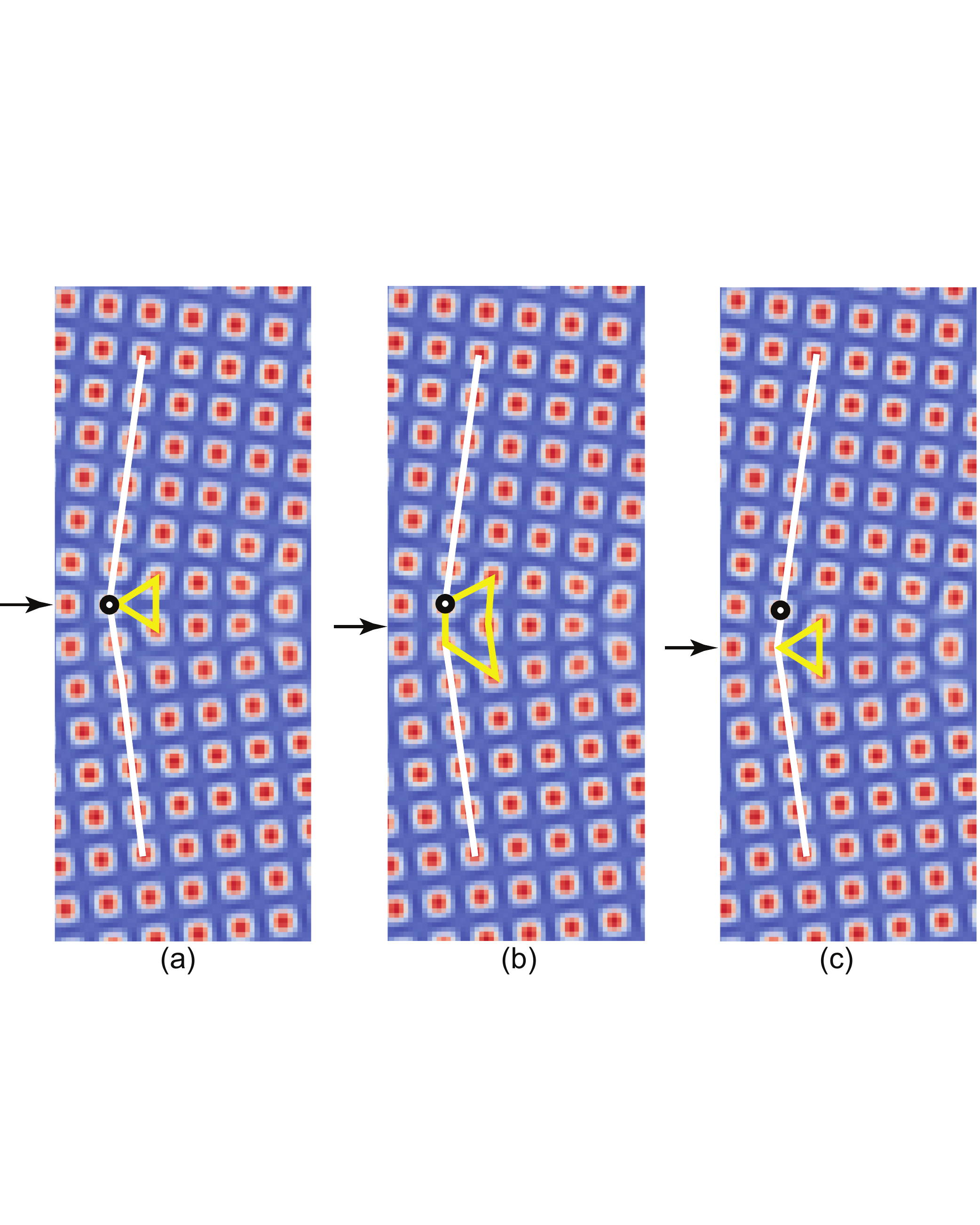} 
\par\end{centering}

\caption{PFC observation of $1/2\langle11\rangle$ dislocation glide during
coupled motion of the symmetrical GB with $\theta=16.26^{\circ}$
and $\phi=45{}^{\circ}$. Frames (a), (b) and (c) show sequential
configurations of the boundary moving down, with the current GB position
shown by an arrow. The dislocation core is marked by a yellow triangle.
The white line connects a selected set of sites (the same in all three
frames) located in a $(11)$ plane. One of the sites is labeled by
an open circle for tracking.\label{fig:PFC-disloc-glide-(11)}}
\end{figure}

\begin{figure}
\begin{centering}
\includegraphics[scale=0.8]{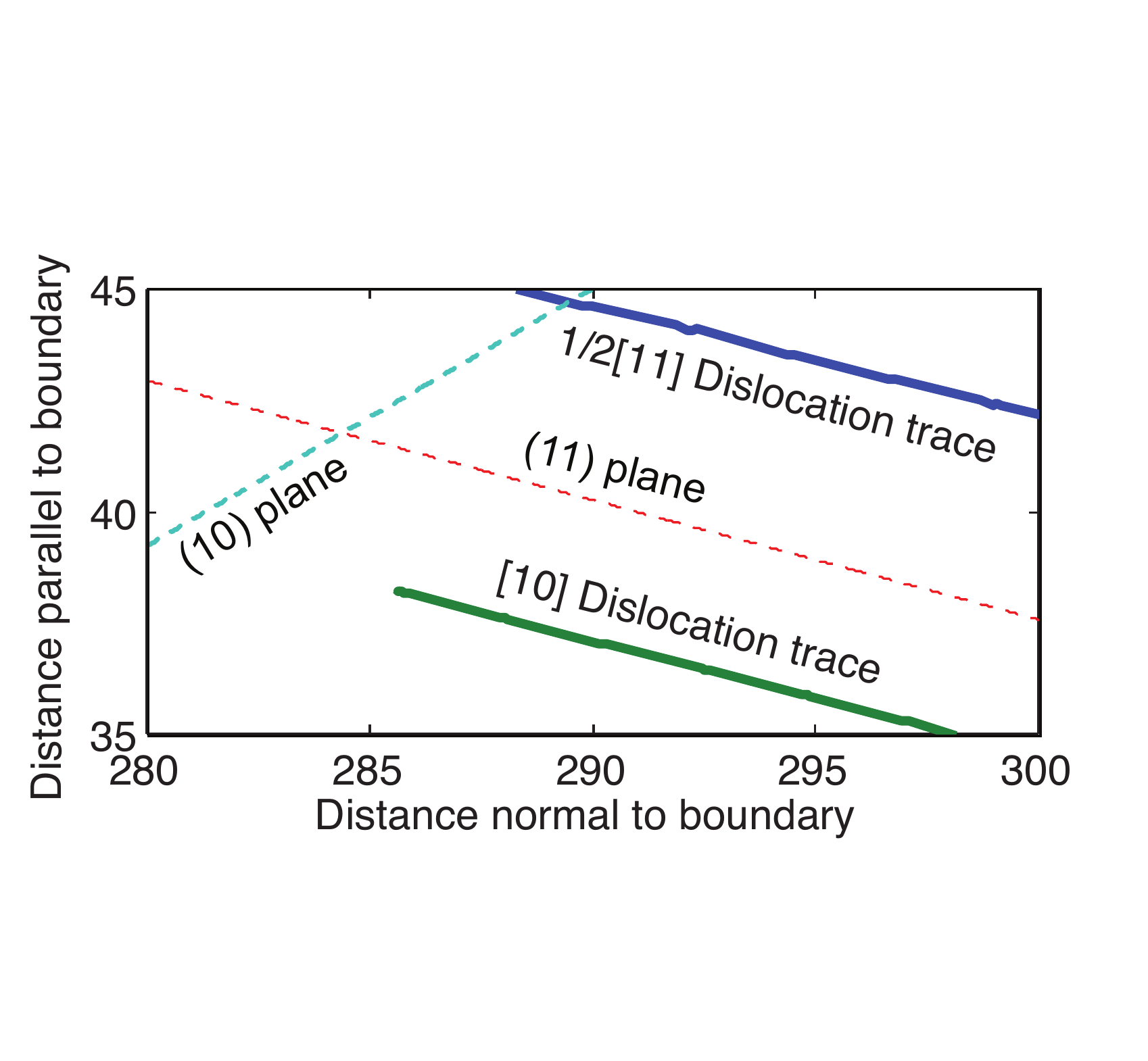} 
\par\end{centering}

\caption{Typical dislocation traces in PFC simulations of coupled motion of
an asymmetrical GB with $\theta=16.26^{\circ}$ and $\phi=30.3{}^{\circ}$.
The dashed lines indicate $(11)$ and $(10)$ crystal planes in the
advancing grain. Both $\langle10\rangle$ and $1/2\langle11\rangle$
dislocations move parallel to $(11)$ planes, suggesting that the
motion of $\langle10\rangle$ dislocations involves climb. The distances
are in Angstroms.\label{fig:PFC-disloc-trace2}}
\end{figure}

\begin{figure}
\begin{centering}
\includegraphics[clip,scale=0.65]{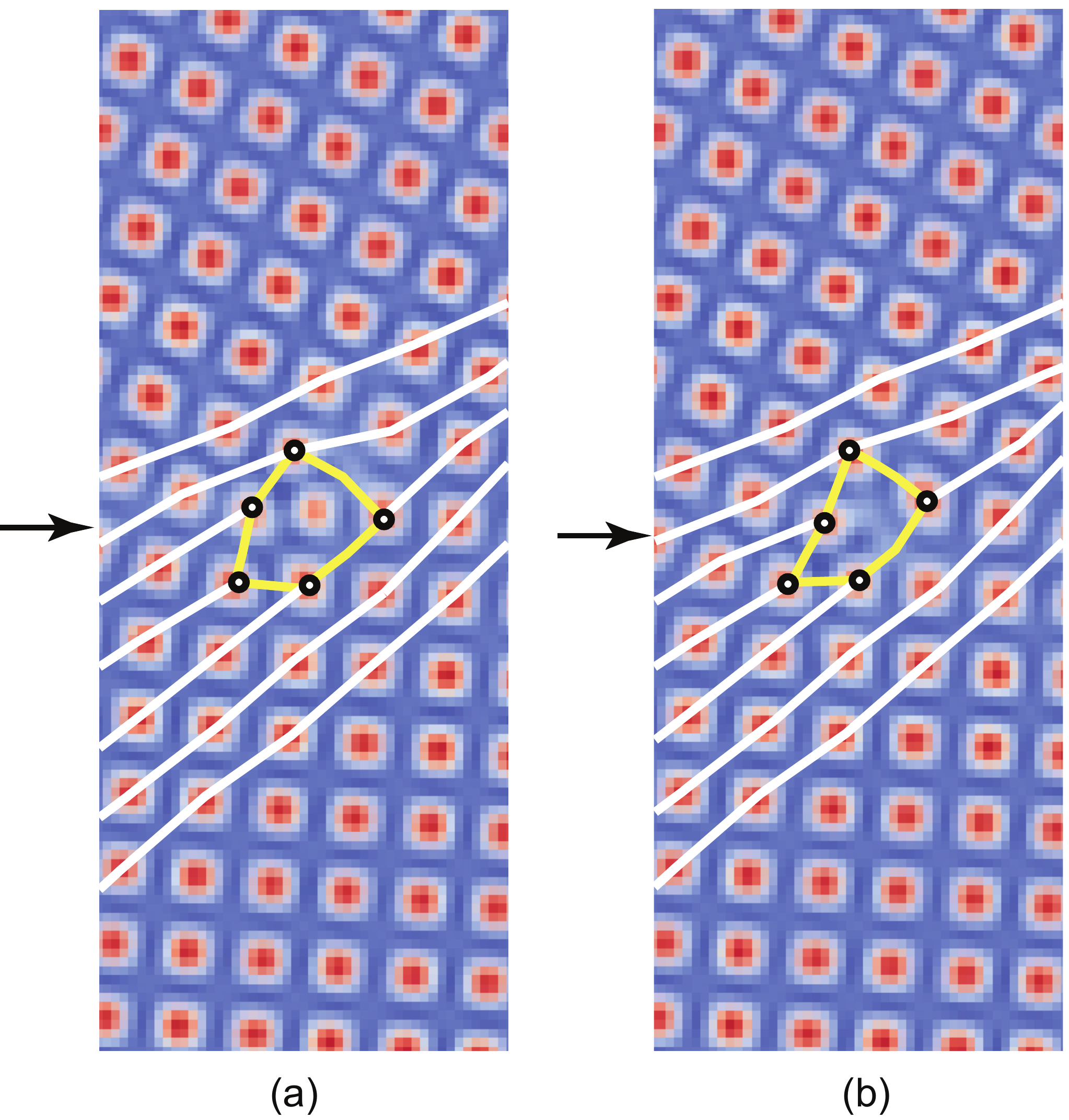} 
\par\end{centering}

\caption{PFC observation of $\langle10\rangle$ dislocation motion during coupled
migration of an asymmetrical GB with $\theta=16.26^{\circ}$ and $\phi=30.3{}^{\circ}$.
Frames (a) and (b) show sequential configurations of the boundary
moving down, with the current GB position shown by an arrow. The dislocation
core region is outlined by five selected sites marked by open circles
and connected by yellow lines. In (a), the yellow polygon encircles
a sixth site, whereas in (b) this site disappears, providing evidence
of dislocation climb. The white lines outline some of the $(20)$
planes as a guide to the eye.\label{fig:PFC-climb} }
\end{figure}

\newpage{}\clearpage{}

\begin{figure}
\noindent \begin{centering}
\includegraphics[clip,scale=0.9]{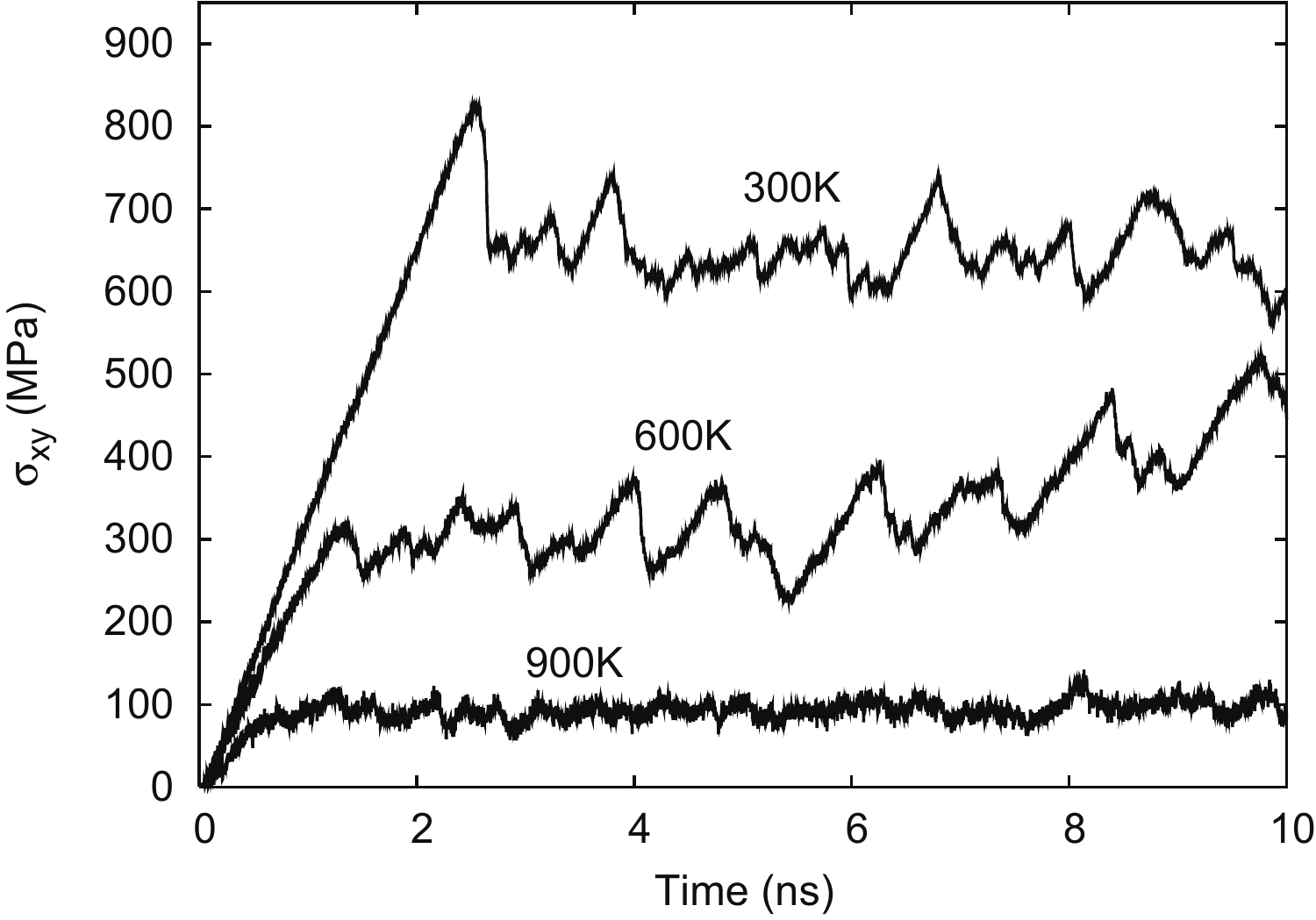} 
\par\end{centering}

\caption{Typical time dependencies of the shear stress during coupled motion
of an asymmetrical GB in Al with $\theta=16.26^{\circ}$ and $\phi=-18.44{}^{\circ}$
at three different temperatures. \label{fig:stress-time}}
\end{figure}

\begin{figure}
\begin{centering}
\textbf{\Large (a)}{\Large{} \includegraphics{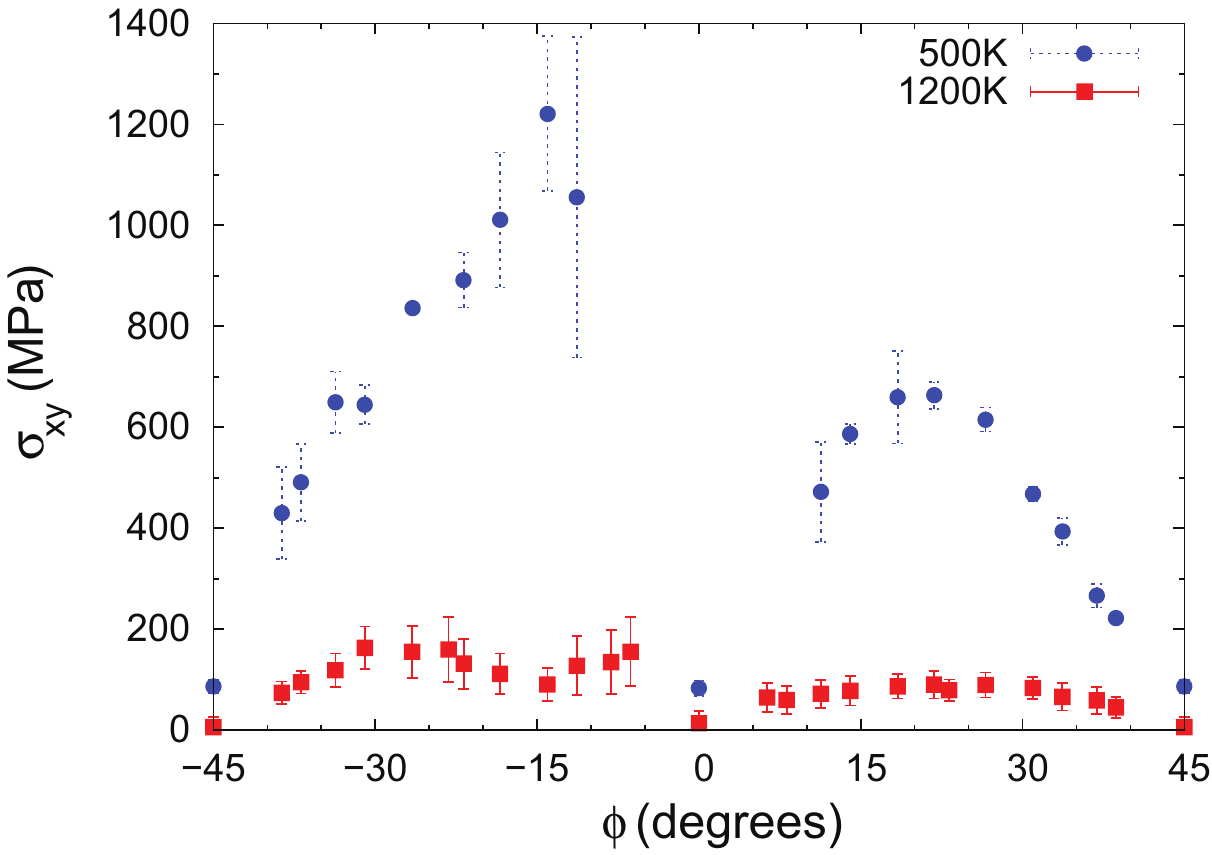} }
\par\end{centering}{\Large \par}

\begin{centering}
\bigskip{}

\par\end{centering}

\begin{centering}
\bigskip{}

\par\end{centering}

\begin{centering}
\textbf{\Large (b)}{\Large{} \includegraphics[clip]{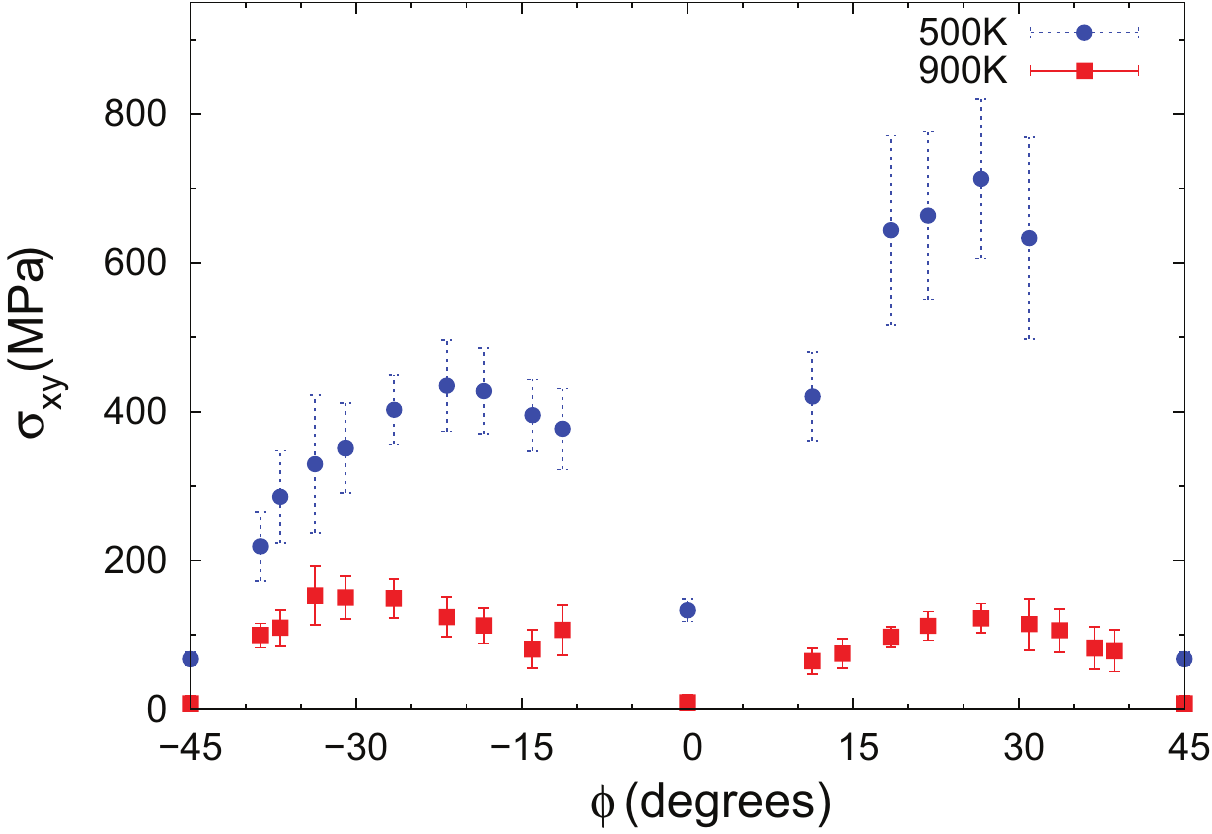} }
\par\end{centering}{\Large \par}

\caption{Steady-state shear stress as a function of inclination angle for GBs
with the tilt angle $\theta=16.26^{\circ}$. (a) Results for Cu. (b)
Results for Al. The simulation temperatures are indicated in the legends.\label{fig:stress-angle} }
\end{figure}

\begin{figure}
\begin{centering}
\textbf{\Large (a)}{\Large{} \includegraphics{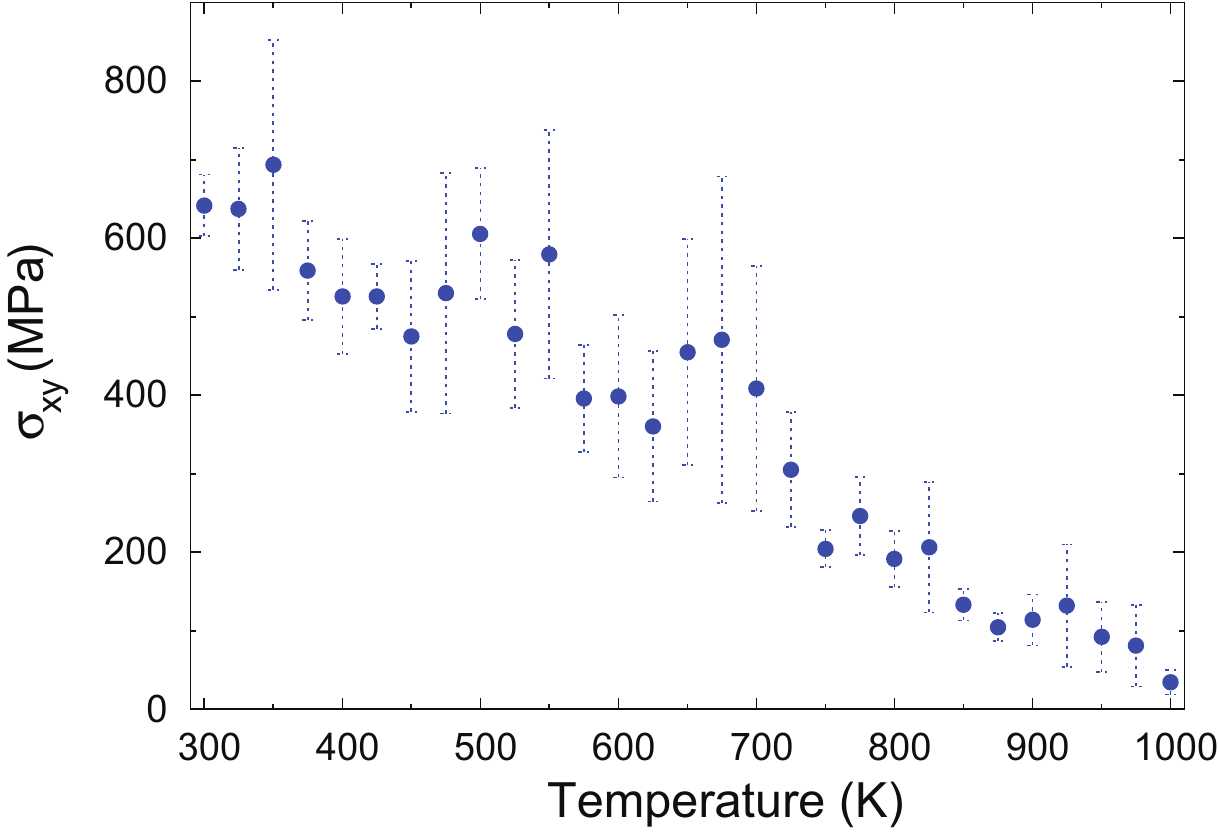} \bigskip{}
}
\par\end{centering}{\Large \par}

\begin{centering}
\bigskip{}

\par\end{centering}

\begin{centering}
\textbf{\Large (b)}{\Large{} \includegraphics{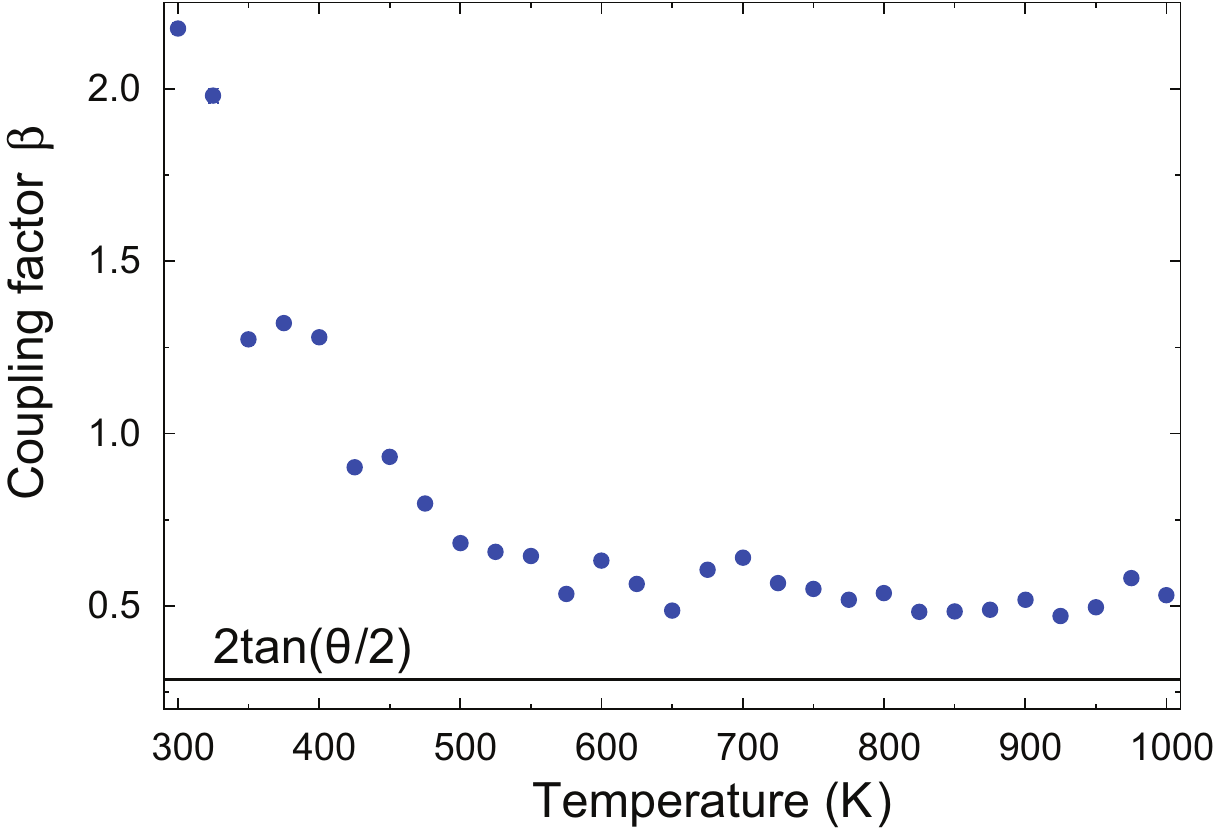} }
\par\end{centering}{\Large \par}

\caption{Shear stress (a) and coupling factor (b) as functions of temperature
for an Al GB with $\theta=16.26^{\circ}$ and $\phi=-18.44^{\circ}$.
In (b), the horizontal line indicates the ideal coupling factor. \label{fig:Coupling-vs-Temperature}}
\end{figure}

\begin{figure}
\noindent \begin{centering}
\includegraphics{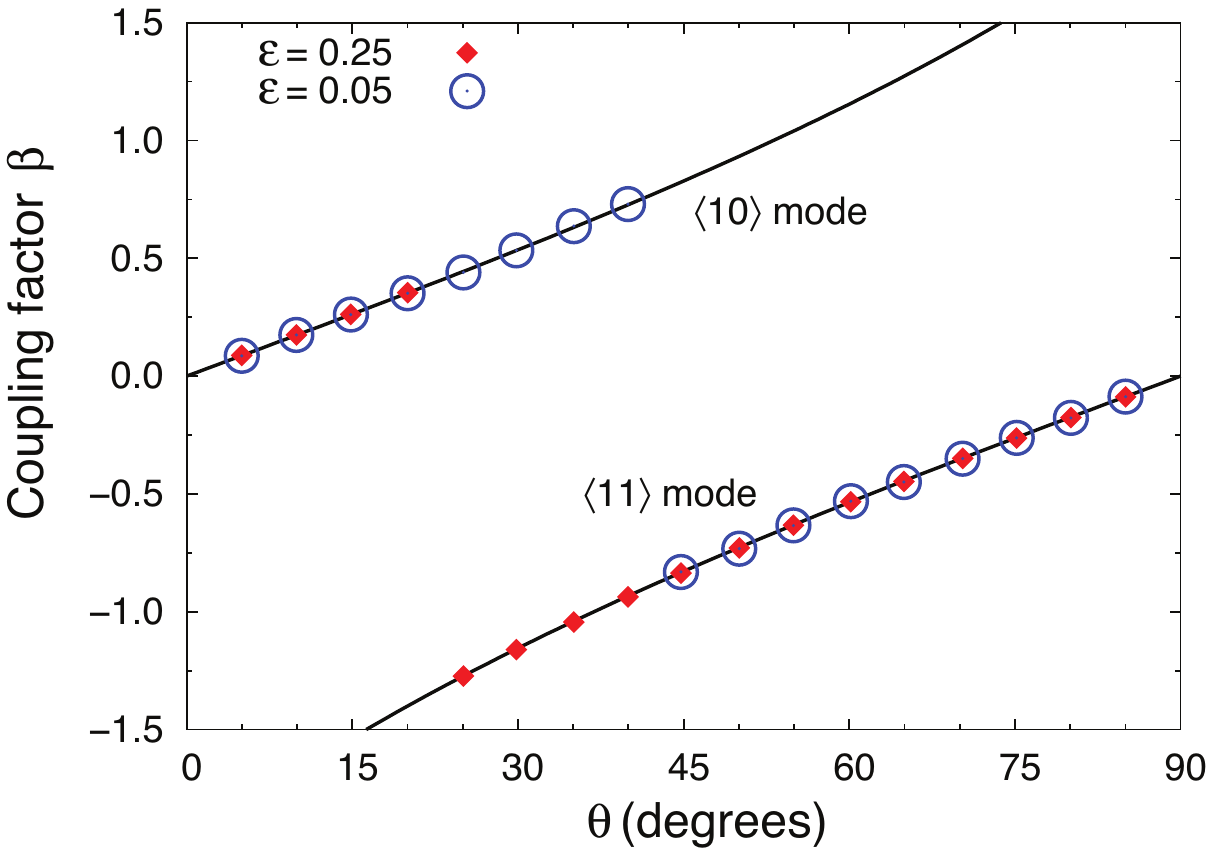} 
\par\end{centering}

\caption{Coupling factor of symmetrical tilt GBs as a function of misorientation
angle obtained by PFC simulations with $\epsilon=0.25$ and $\epsilon=0.05$.
The lines indicate theoretical predictions \citep{Cahn06b} for two
coupling modes.\label{fig:beta-theta} }
\end{figure}

\begin{figure}
\begin{centering}
\textbf{\Large (a)}{\Large{} \includegraphics{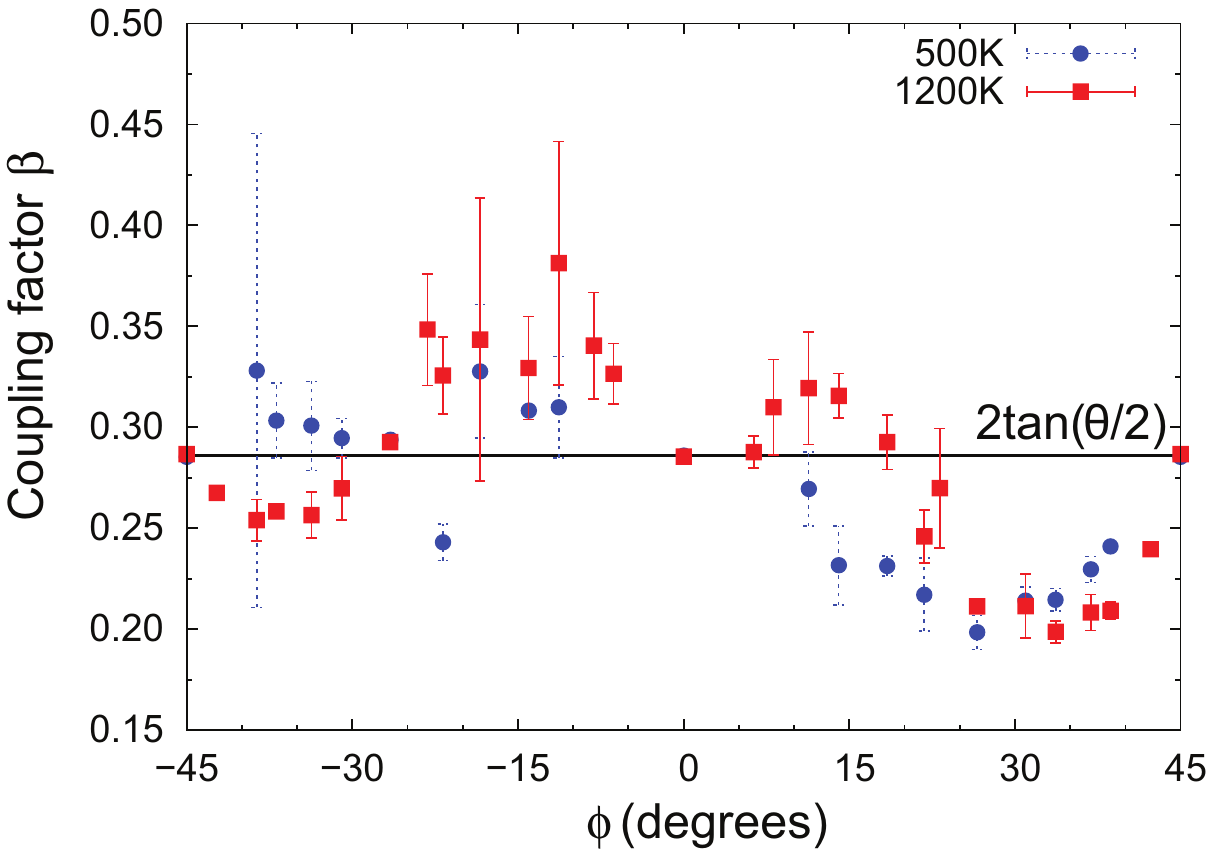} }
\par\end{centering}{\Large \par}

\begin{centering}
\bigskip{}

\par\end{centering}

\begin{centering}
\bigskip{}

\par\end{centering}

\begin{centering}
\textbf{\Large (b)}{\Large{} \includegraphics{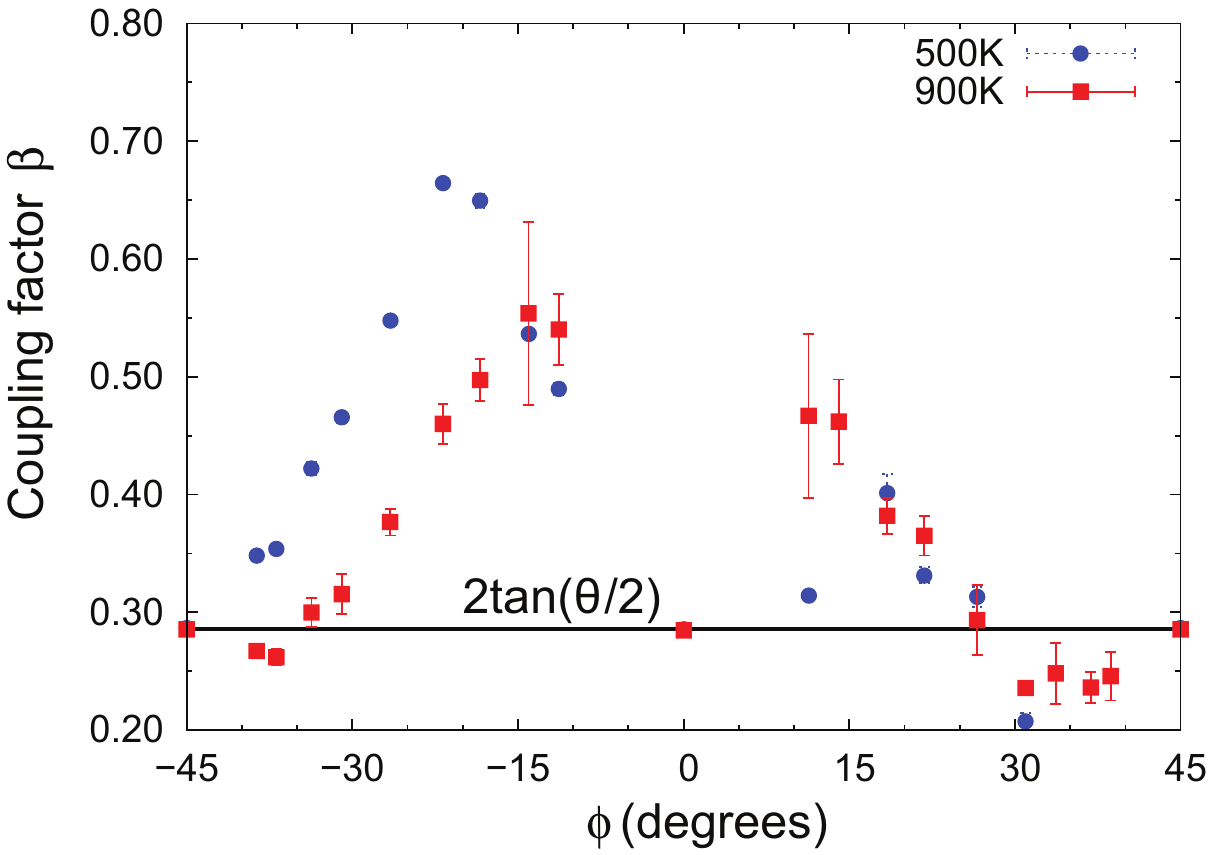} }
\par\end{centering}{\Large \par}

\caption{MD simulation results for the coupling factor of asymmetrical GBs
in (a) Cu and (b) Al with $\theta=16.26^{\circ}$ as a function of
the inclination angle $\phi$. The simulation temperatures are indicated
in the legends. The horizontal line indicates the ideal coupling factor.\label{fig:beta-Cu-16}}
\end{figure}

\begin{figure}
\noindent \begin{centering}
\includegraphics{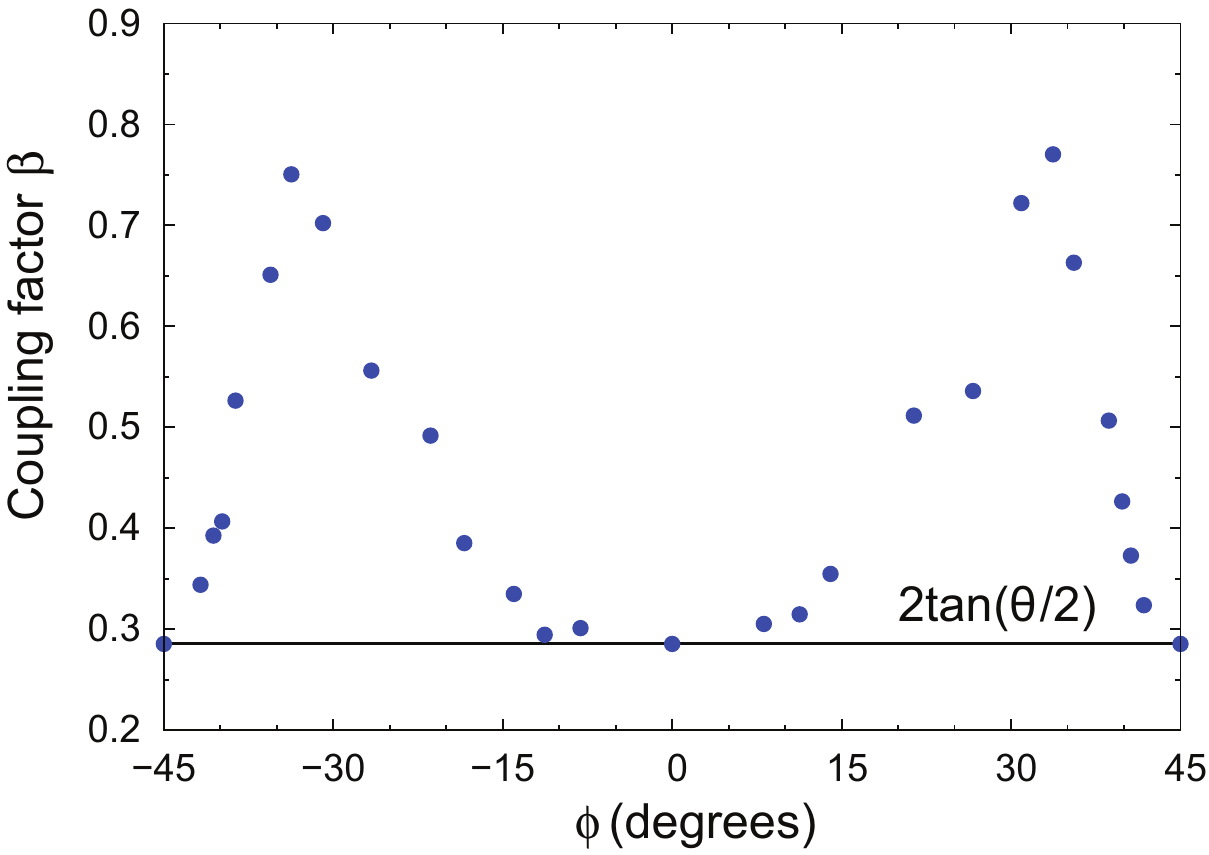} 
\par\end{centering}

\caption{PFC results for the coupling factor of asymmetrical GBs with $\theta=16.26^{\circ}$
as a function of the inclination angle $\phi$. The horizontal line
indicates the ideal coupling factor.\label{fig:PFC-beta-16}}
\end{figure}

\begin{figure}
\begin{centering}
\textbf{\Large (a)}{\Large{} \includegraphics[scale=0.95]{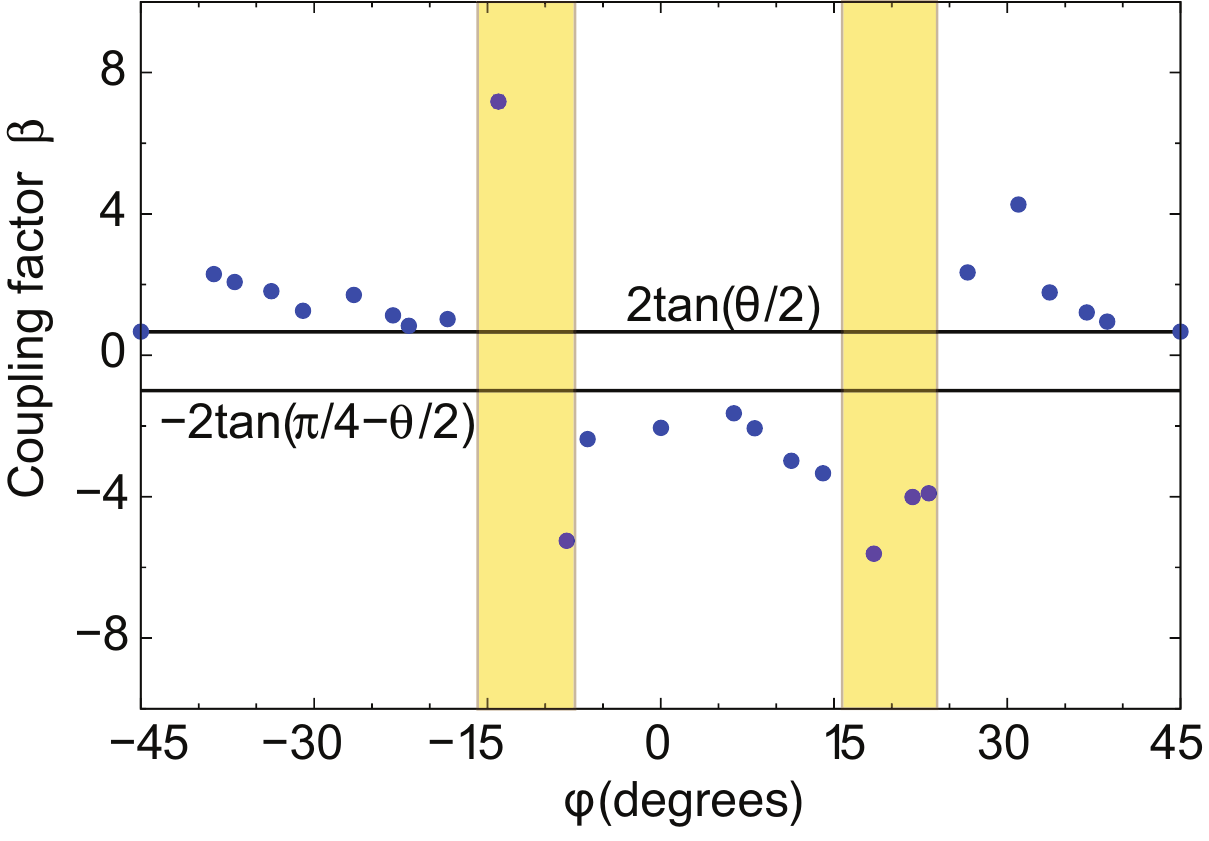} }
\par\end{centering}{\Large \par}

\begin{centering}
\bigskip{}

\par\end{centering}

\begin{centering}
\bigskip{}

\par\end{centering}

\begin{centering}
\textbf{\Large (b)}{\Large{} \includegraphics[scale=0.95]{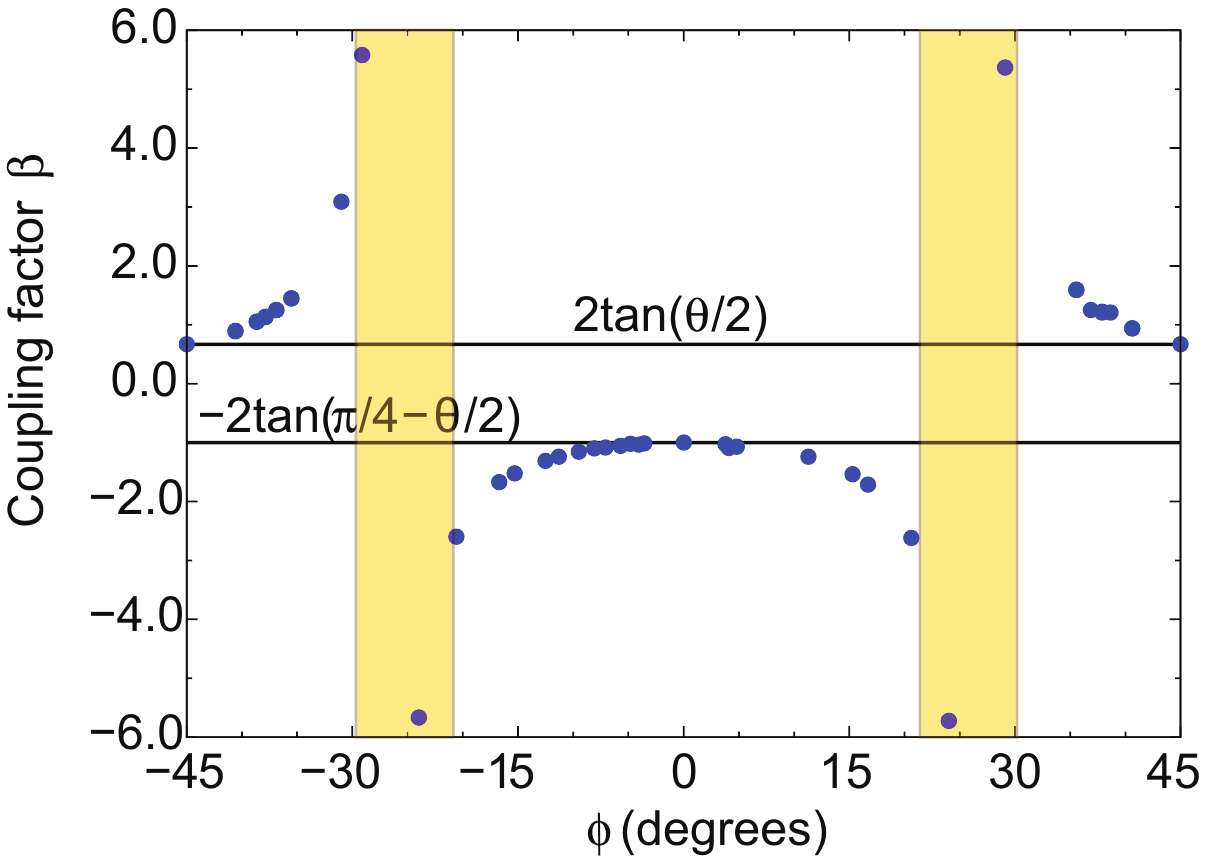} }
\par\end{centering}{\Large \par}

\caption{MD simulation results for Cu at 800 K (a) and PFC simulation results
(b) for the coupling factor of asymmetrical GBs with $\theta=36.87^{\circ}$
as a function of the inclination angle $\phi$. The horizontal lines
indicate the ideal coupling factors for two coupling modes. The shaded
stripes indicate approximate regions in which the coupling factor
switches between the modes. \label{fig:beta-Cu-37}}
\end{figure}

\begin{figure}
\noindent \begin{centering}
\includegraphics{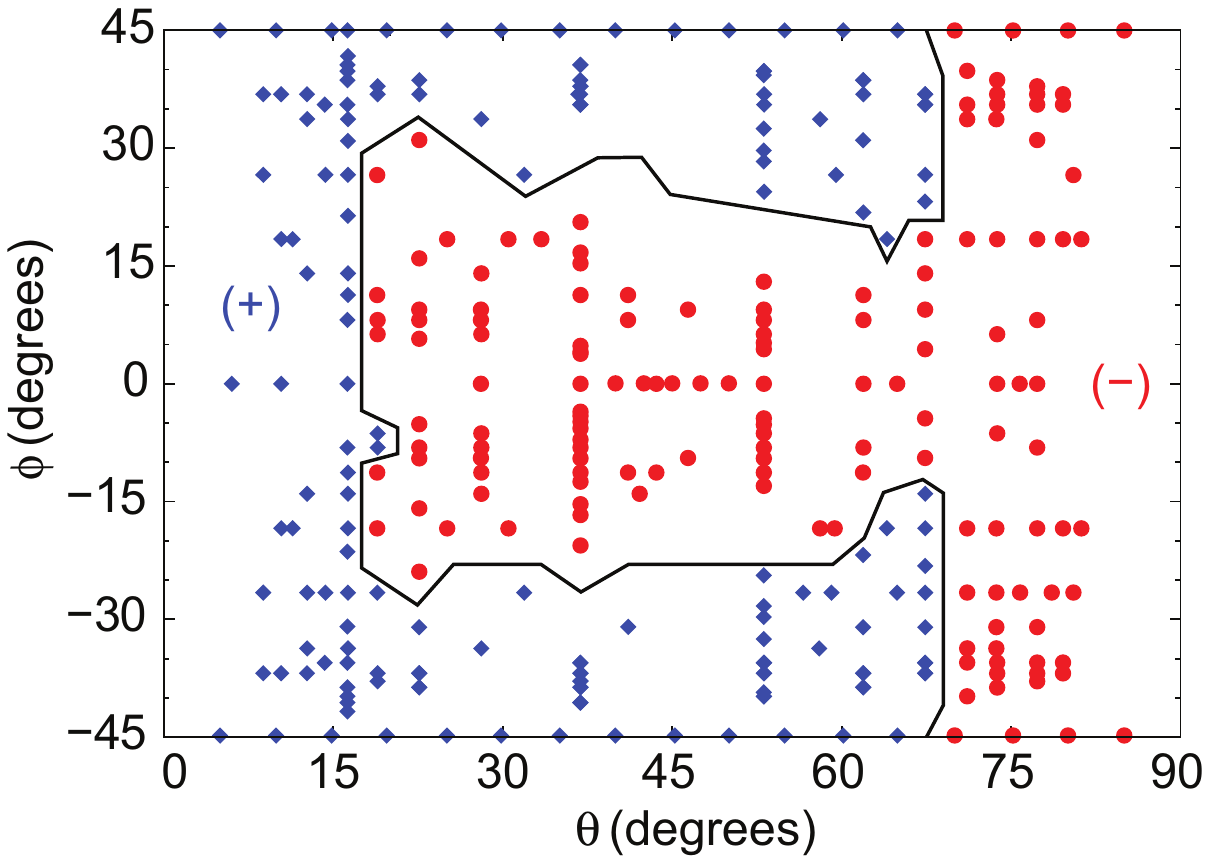} 
\par\end{centering}

\caption{Summary of PFC calculations of the coupling factor for asymmetrical
GBs. The diamond and circle symbols indicate positive and negative
$\beta$, respectively. The line outlines the approximate boundary
between the two coupling modes with different signs of $\beta$.\label{fig:Summary-of-PFC}}
\end{figure}

\end{document}